\documentclass[sn-mathphys-num]{sn-jnl}

\usepackage{graphicx}%
\usepackage{multirow}%
\usepackage{amsmath,amssymb,amsfonts}%
\usepackage{amsthm}%
\usepackage{mathrsfs}%
\usepackage[title]{appendix}%
\usepackage{xcolor}%
\usepackage{textcomp}%
\usepackage{manyfoot}%
\usepackage{booktabs}%
\usepackage{algorithm}%
\usepackage{algorithmicx}%
\usepackage{algpseudocode}%
\usepackage{listings}%
\usepackage{tabularx}
\usepackage{booktabs}
\usepackage{adjustbox}
\usepackage{hyperref}

\theoremstyle{thmstyleone}%
\theoremstyle{thmstyletwo}%

\theoremstyle{thmstylethree}%

\begin{document}

\title[Article Title]{Building Intelligence Identification System via Large Language Model Watermarking: A Survey and Beyond}


\author[1]{\fnm{Xuhong} \sur{Wang}}\email{wangxuhong@pjlab.org.cn}
\equalcont{These authors contributed equally to this work.}

\author[2,1]{\fnm{Haoyu} \sur{Jiang}}\email{jhy549@sjtu.edu.cn}
\equalcont{These authors contributed equally to this work.}
\author[1]{\fnm{Yi} \sur{Yu}}\email{yuyi@pjlab.org.cn}
\author[1]{\fnm{Jingru} \sur{Yu}}\email{yujingru@pjlab.org.cn}
\author*[1]{\fnm{Yilun} \sur{Lin}}\email{linyilun@pjlab.org.cn}
\author*[2]{\fnm{Ping} \sur{Yi}}\email{yiping@sjtu.edu.cn}
\author[1]{\fnm{Yingchun} \sur{Wang}}\email{wangyingchun@pjlab.org.cn}
\author[1]{\fnm{Yu} \sur{Qiao}}\email{qiaoyu@pjlab.org.cn}
\author[3]{\fnm{Li} \sur{Li}}\email{li-li@mail.tsinghua.edu.cn}
\author[4]{\fnm{Fei-Yue} \sur{Wang}}\email{feiyue.wang@ia.ac.cn}

\affil[1]{\orgname{Shanghai Artificial Intelligence Laboratory}, \orgaddress{ \postcode{200433}, \state{Shanghai}, \country{China}}}

\affil[2]{
\orgdiv{School of Cyber Science and Engineering},
\orgname{Shanghai Jiao Tong University}, \orgaddress{ \postcode{200240}, \state{Shanghai}, \country{China}}}

\affil[3]{
\orgdiv{Department of Automation, BNRist},
\orgname{Tsinghua University}, \orgaddress{ \postcode{100084}, \state{Beijing}, \country{China}}}

\affil[4]{
\orgdiv{The State Key Laboratory for Management and Control of Complex Systems, Institute of Automation},
\orgname{Chinese Academy of Sciences}, \orgaddress{ \postcode{100190}, \state{Beijing}, \country{China}}}


\abstract{Large Language Models (LLMs) are increasingly integrated into diverse industries, posing substantial security risks due to unauthorized replication and misuse. To mitigate these concerns, robust identification mechanisms are widely acknowledged as an effective strategy. Identification systems for LLMs now rely heavily on watermarking technology to manage and protect intellectual property and ensure data security. However, previous studies have primarily concentrated on the basic principles of algorithms and lacked a comprehensive analysis of watermarking theory and practice from the perspective of intelligent identification. To bridge this gap, firstly, we explore how a robust identity recognition system can be effectively implemented and managed within LLMs by various participants using watermarking technology. Secondly, we propose a mathematical framework based on mutual information theory, which systematizes the identification process to achieve more precise and customized watermarking. Additionally, we present a comprehensive evaluation of performance metrics for LLM watermarking, reflecting participant preferences and advancing discussions on its identification applications. Lastly, we outline the existing challenges in current watermarking technologies and theoretical frameworks, and provide directional guidance to address these challenges. Our systematic classification and detailed exposition aim to enhance the comparison and evaluation of various methods, fostering further research and development toward a transparent, secure, and equitable LLM ecosystem.
}




\maketitle

\section{Introduction}\label{sec1}
Large Language Models  (LLMs) have become increasingly important for driving innovation across multiple industries. From automated customer service to complex natural language understanding tasks, the applications of LLMs are expanding. However, as LLMs become more widely used, the challenges to protect security, compliance, and user privacy have become increasingly severe, highlighting the urgent need for robust identity recognition systems.

Identity recognition plays a crucial role across various sectors in modern society~\cite{jain2004introduction}, from financial transactions~\cite{prabakaran2022multi} and healthcare~\cite{ren2014user} to border security~\cite{labati2016biometric} and online services~\cite{talreja2018biometrics}. The application of identity recognition technology is ubiquitous, ensuring user authentication and authorization, and serving as the cornerstone of security and privacy. In fact, all existing governance frameworks and security systems rely on the effective operation of identity recognition systems~\cite{grassi2020digital}. Despite the widespread application of identity recognition systems in many fields, such systems have yet to be fully established in the realm of artificial intelligence (AI). This is primarily due to the complexity and dynamic nature of the AI domain, where traditional identity recognition methods struggle to meet the demands of AI systems. The core issues of identity recognition involve achieving distinguishability, unforgeability, and traceability. These issues are particularly critical in the context of LLMs, where the characteristics of textual data, the openness of LLMs, and the extensive applications of LLMs make identity recognition even more complex.


Currently, watermarking technology is regarded as a potential solution to address the three core issues in identity recognition~\cite{zhu2005survey}. It can covertly embed identity information without compromising the quality of the original data~\cite{kamaruddinReviewTextWatermarking2018}, ensuring distinguishability. By integrating cryptography, watermarking technology ensures the unforgeability of information and enables traceability through detection. This technology offers an innovative strategy for intellectual property protection and data security in the field of LLMs. Given the urgent need to protect intellectual property and ensure the traceability of security responsibilities in complex LLM application scenarios, it is essential to establish effective techniques and theoretical frameworks for embedding and extracting watermarks.


Although some existing literature reviews~\cite{yangSurveyDetectionLLMsGenerated2023,wuSurveyLLMGeneratedText2024,liuSurveyTextWatermarking2024,yaoSurveyLargeLanguage2024} have gradually focused on these issues, most studies primarily introduce the basic principles of algorithms. They lack a comprehensive analysis of watermarking as the cornerstone of identity recognition systems for LLMs and do not adequately address the multifaceted conflicts of interest encountered by LLMs in actual operation.
This article innovates the existing LLM watermarking systems from three main aspects: application, theory, and evaluation, thereby providing theoretical and practical support for the secure, transparent, and fair use of LLMs. The main contributions of this article are as follows.

\textbf{Application:} In Section~\ref{sec2}, we illustrate that the LLM application system is transitioning from a centralized setup, dominated by model technology service providers, to a multi-centric design that prioritizes identity verification and behavior traceability.
We also explored the different preferences of data providers, technology service providers, users, and third-party regulators regarding various aspects of identity recognition systems within a multi-center LLM application framework.
This novel perspective deepens our understanding of the rights and responsibilities of participants in LLM community. It also promotes the establishment of a fairer and more secure AI application environment.

\textbf{Theory:} In Section~\ref{sec3}, we address the limitations of current LLM watermarking technology by developing a theoretical system based on mutual information theory~\cite{cover1999elements}. The comprehensive mathematical foundation establishes a formulaic framework and classifies LLM watermarking technologies into five primary processes: generation, embedding, attack, extraction, and reconstruction. The optimization object and constraints of each process are elaborated with mathematical formulas, allowing researchers to accurately develop and enhance the watermarking techniques based on corresponding roles and stages. 


\textbf{Evaluation:} In Section~\ref{sec4}, we have synthesized the performance evaluation metrics for LLM watermarks from multiple perspectives, encapsulating the preferences of various LLM entities in their application of watermarking techniques for identity recognition. This summary contributes to the development of a comprehensive and standardized evaluation system, prompting consideration of security issues related to LLM watermarking, and outlines new research trajectories and technological orientations.

Through these three core contributions, our article significantly expands the scope of watermarking applications within LLMs. For the first time, we put watermarking techniques within the context of the identification applications of LLMs, providing robust technical support for addressing the challenges of security and transparency in LLMs. This integration serves a dual purpose: it enhances the traceability of content generated by LLMs, allowing each output to be reliably traced back to its originating model, and it substantially boosts the trustworthiness of LLMs in various application scenarios by ensuring the authenticity and provenance of the content. Finally, we have highlighted some challenges that still exist in the current watermarking technology and theoretical systems, and suggested potential solutions for these challenges.
We hope this work will spark further research and discussion, propelling LLM technology towards a trustworthy and verifiable future while safeguarding user interests.

\section{Establishing Identification System through LLM Watermarking}\label{sec2}
\subsection{Future Trends in LLM Applications}
Currently, the research and development of LLMs are in a period of rapid growth, benefiting from the swift enhancement of computing power and the accumulation of large volumes of high-quality data. The life cycle of LLMs can generally be divided into stages of data preparation, training and testing, deployment and application, and monitoring and maintenance.  Entities and participants typically involved in the life cycle of LLMs include training data providers (such as Stardust AI\footnote{\url{https://stardust.ai/en-US/}}, Scale AI\footnote{\url{https://scale.com/}}, etc.), model technology service providers (such as OpenAI\footnote{\url{https://openai.com/}}, Anthropic\footnote{\url{https://www.anthropic.com/}}, etc.), LLM users, and certain public
regulators and trusted third parties (PRTTPs)  (governments, non-profit organizations, etc.)\footnote{It is important to note that not all entities are involved in the research and development of every LLM. For example, certain companies dedicated to LLMs handle data collection and cleaning internally, and some models, remaining closed to the public, consequently do not require regulation.}. 

\begin{figure}[t]
    \centering
    \includegraphics[width=\linewidth]{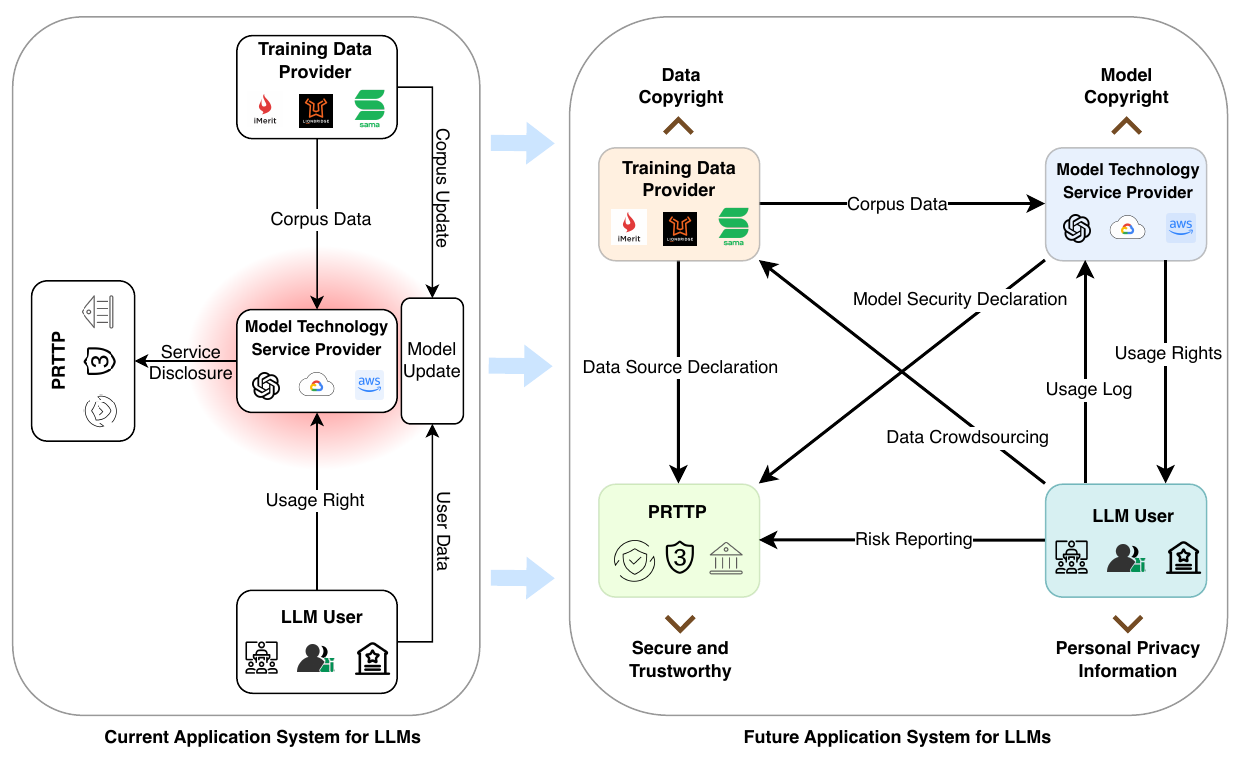}
    \caption{Evolution of application systems for LLMs: transitioning from a centralized system focused on model technology service providers to a multi-centric system emphasizing identity verification and behavior traceability.}
    \label{fig:1}
\end{figure}

However, people tend to focus on the iteration of technology (models, algorithms, and data) while neglecting issues of security and rights protection in the application processes of LLMs.
As shown in the left part of Fig.~\ref{fig:1}, in the existing LLM R\&D system, technology service providers play a dominant role in every step, from data preparation and model training to final deployment and maintenance, relying on their technical reserves and commercial needs. This centralized system allows technology service providers to monopolize the entire LLM technology market through technological barriers and resource advantages, making it difficult for other participants to develop or achieve breakthroughs independently. Overall, this system makes the development of technology, model compliance, and user privacy security dependent on the ethical standards of technology service providers, which is not conducive to the overall development of the AI ecosystem. 

Typically, once the LLM technologies have been widely promoted and enter a period of stabilization, the LLM community shifts its focus from solely valuing the technology to emphasizing regulatory compliance, user privacy, and the security of the technology. This shift transforms the original technology-centric centralized operational system into a balanced, multi-centric system involving multiple participants, as illustrated in the right part of Fig.~\ref{fig:1}. In this system, the influence of users and PRTTPs is significantly enhanced. PRTTPs are responsible for obtaining and verifying security and trust declarations from data providers and technology service providers, as well as handling risk reports from general LLM users. Meanwhile, LLM users, while ensuring their security and privacy, will authorize the collection of their preferences to model technology service providers and gain potential benefits.

In this new system, as the status of each entity becomes more balanced, they all seek to maximize their benefits while ensuring their rights are protected, rather than merely using the model passively. For instance, LLM users might suspect that the model technology service providers could steal their privacy; data providers might worry that service providers could resell their data. The most critical aspect of ensuring the flawless operation of the entire lifecycle system of LLMs is to ensure that these entities can engage in trustworthy collaboration through certain mechanisms, thereby minimizing mutual suspicion to the greatest extent. The core element in reducing suspicion is making the identities in the LLM recognizable and their behaviors traceable. 


\subsection{Identity Recognition System in LLMs}
In the current digital era, identity recognition technology is critical for safeguarding information security. Traditional identity recognition techniques, such as Multi-Factor Authentication (MFA)~\cite{ometov2018multi}, biometric technologies~\cite{jain2000biometric} (including fingerprint~\cite{bebis1999fingerprint} and facial recognition~\cite{kaur2020facial}), and Single Sign-On (SSO)~\cite{de2002single}, are primarily employed to authorize and identify individuals within human communities, relying on biometric features, behavioral patterns, and language analysis to verify identities. For instance, some studies identify individuals by modeling their interaction behaviors with devices~\cite{shen2017pattern} and their language styles~\cite{abbasi2022authorship}. However, an effective identity recognition system has yet to emerge in the LLM community.


Due to the capacity of LLMs to generate text of high quality and diversity, it poses a novel challenge to authenticate whether a segment of text is the output of a particular LLM, and to confirm that it has not been unauthorizedly altered or counterfeited. Traditional identity recognition technologies are not applicable in this scenario, as they cannot be directly implemented on the text content or LLMs. 


Watermarking technology is a crucial method for identity recognition in the field of computer science~\cite{saini2014survey,li2019prove,ahvanooey2020anitw,boenisch2021systematic}. Traditionally used for copyright protection in images, audio, and video~\cite{potdarSurveyDigitalImage2005}. Watermarking embeds secret information without compromising original data quality. With the rise of LLM technology, embedding watermark information in LLMs themselves and their related applications has become an indispensable area of research. Watermarking enables verification of text origin from specific LLMs, thereby enhancing copyright protection, intellectual property preservation, and content authenticity. Moreover, watermarking aids in tracing content dissemination, preventing misinformation, and ensuring transparency and traceability in compliance with legal and regulatory standards. Consequently, watermarking technology has emerged as an innovative and indispensable mechanism for identification in the context of LLM applications.

\subsection{Identification System from Different Views}


In practical application scenarios, data providers can use watermarks to protect the copyright of their training data, ensuring that the training data are not arbitrarily altered or copied. Technology service providers wish to use watermarking technology to protect their model copyrights, preventing their models from being repackaged or stolen, and enabling them to track the usage of their models. LLM users need watermarking technology to protect their privacy rights, preventing their confidential information from being sold. PRTTP will ensure the security of LLMs by verifying the presence of watermarks at multiple stages; once any security issues are identified, it is crucial to ensure that the source of the problem can be traced and resolved. The following sections detail the four distinct entities of the watermark system and elucidate how each can establish its own watermarking technology framework.

\subsubsection{Training Data Providers}
For training data providers, the infinite replicability of data poses an uncontrollable risk of data breaches as it circulates. Currently, the most effective way to prevent the unauthorized dissemination of data is to secretly add a unique, strong watermark~\cite{liFunctionMarkerWatermarkingLanguage2023} to the data without altering the quality of the dataset itself~\cite{liuWatermarkingClassificationDataset2023a}. This ensures that the data can still be verified for its initial copyright even after being redistributed, modified, or otherwise processed. Some researchers have proposed even more in-depth solutions, demonstrating that watermarked data used for LLM training can have its watermark information detected in the text generated by the LLM, which is referred to as radioactivity~\cite{sanderWatermarkingMakesLanguage2024a}. Once this knowledge is embedded into unauthorized LLMs, data providers can identify whether their watermark is present in the models based on the response to certain specific watermark triggers. Moreover, since a training dataset is likely to be copied and sold multiple times, the most crucial aspect of protecting copyright and preventing data leaks or unauthorized distribution is identifying the source of the leakage. To address this, encoding a unique watermark message for each dataset to be distributed and embedding it into the data with a covert watermark is an essential option.

Therefore, training data providers need to focus on watermarking techniques that offer high fidelity and transparency, meaning the watermarking should be done in a way that does not degrade the quality of the text. Moreover, the watermark embedded by the data providers should have the capability of multi-bit information encoding to enable the identification of the data purchasers. Besides, the watermark should possess high robustness and radioactivity, allowing for the detection of watermarks in text content generated by unauthorized models if the data are used for illegal training.

\subsubsection{Model Technology Service Providers}
For technology service providers, watermarking technology helps protect model copyrights and monitor the usage of models. To address the costs and technical difficulties faced during the pre-training of LLMs, some technology service providers might opt to use data generated by well-trained models for training, which has formed a system similar to teacher-student model distillation. This imitation has sparked concerns over the copyright of unauthorized distilled models, especially when the corpus data of these distilled models come from closed-source LLMs (such as GPT-4). In the constantly evolving landscape of AI copyright protection, it is crucial to emphasize the importance of protecting intellectual property while maintaining the integrity and practicality of AI models. The development and implementation of watermarking technology enables model developers to protect their innovations from unauthorized use and distribution effectively.

To protect the intellectual property of models and prevent the unauthorized use of developed LLMs through distillation by offenders, technology service providers should embed watermarks only when the model is invoked by users, without affecting the model's own training process. This approach meets the technology service providers' pursuit of model performance. Additionally, it substantiates the model's ownership and facilitates the tracking of its distribution and usage~\cite{zhang2021deep}. This helps prevent the model from being copied or tampered with by unauthorized third parties.

\subsubsection{Public Regulators and Trusted Third Parties}
With the rapid advancement of AI technology, especially the widespread use of LLMs in content creation, the roles of public regulators and trusted third parties (PRTTPs) in watermarking systems have become critical. Policymakers and civil society are increasingly focused on the safe use of these technologies, as shown by the EU AI Act, the NDAA for Fiscal Year 2024~\cite{rep.rogersNationalDefenseAuthorization2023}, and voluntary commitments to label AI-generated content. These initiatives emphasize the need for transparency in content sources, including clear marking of watermarks and content origins, as well as guidelines for content certification and watermarking developed by the U.S. Department of Commerce following the AI Executive Order of October 30, 2023~\cite{FACTSHEETPresident}. These guidelines aim to help the public easily identify the authenticity of online information.

LLM watermarking technology is considered key to ensuring the safety, compliance, and ethical integrity of AIGC throughout its entire lifecycle. To this end, PRTTPs should make the following efforts:
\begin{enumerate}
    \item Establishing clear watermarking technical standards and usage norms, setting up certification programs for watermark service providers, and promoting success stories and best practices.
    \item The implementation of education and training programs, especially aimed at enhancing the understanding of watermarking technology among all participants, further reinforces this process. This involves not only educating parties on how to select and utilize watermark services correctly but also emphasizing the importance of adopting these measures to ensure that all involved can effectively use watermarking technology to identify AIGC.
    \item Establishing strict oversight and enforcement mechanisms is equally important, ensuring that all parties rigorously adhere to the regulations and standards for watermark usage, thereby guaranteeing the correct and secure application of watermarking technology.
    \item Providing the general public with access to open watermark interfaces for LLMs allows users to add watermarks to their own data or to verify through watermarking whether a piece of data contains their private information. This approach helps track the usage and flow of data, preventing unauthorized dissemination of the data across the internet.
\end{enumerate}

These requirements are not isolated but necessitate multi-party collaboration among data providers, model technology service providers, watermark technology service providers, and PRTTPs. Through a cooperation framework that spans different sectors and industries, we can facilitate information sharing and technological advancement. These measures encourage deep reflection on the safe use of AI technology and lay a solid foundation for maintaining public trust in AIGC. By establishing industry standards, implementing rigorous certification processes and audits, providing comprehensive education and training, enforcing vigilant supervision and execution, promoting best practices, and fostering collaborative multi-party partnerships, we can ensure the safe and responsible development of LLMs and other AI technologies.

\subsubsection{Users of LLMs}
Watermarking technology can help LLM users in verifying the copyright and legitimacy of the models, while also serving as a tool to protect the security of user data and privacy. Users should opt for LLM services verified by PRTTPs. The model providers usually require such verified service providers to apply data watermarking technology to ensure that the input data is used only for the current service and is not accessed or misused by third parties. LLM users can ensure that watermarks are embedded in their prompts by choosing services with publicly verifiable watermarking technology. Users can verify the use and flow of their data through a public watermark verification interface, preventing unauthorized distribution of their data on the internet. Watermarking technology plays a crucial role in protecting personal privacy and can also alleviate users' privacy and security concerns on another level, thereby promoting the further popularization of LLM technology.

Moreover, LLM users will gradually transition from mere users to becoming a more deeply involved and crucial part of the LLM application ecosystem. Firstly, users can trade their private data through some form of anonymization, collaborating with data providers to co-create datasets. This not only generates profit but also enhances the overall efficiency of the LLM application system. Secondly, users will engage in in-depth cooperation with PRTTPs. If the model produces unsafe answers or if copyright infringement is detected, users can report these issues to PRTTPs by clicking the report button. PRTTPs can thus centralize the originally dispersed and unequal user oversight power through this method, better standardizing the development of LLM technology.

\subsection{Guidance for Implement Watermarking}
All entities should establish their own rights protection system using watermarking technology based on their position within the LLM application ecosystem and their relationships with other entities. The rights that need protection, the entities that need identification, the limitations encountered when using watermarking technology, and some basic requirements are organized in Table~\ref{table1}. Based on the descriptions in the table, entities can find the corresponding watermarking technologies in Section~\ref{sec3} and deploy their watermarking schemes according to the different basic requirements.

\begin{table}[t]
\caption{Summary of requirements for various entities to implement watermarking schemes in LLMs.}\label{table1}%
\begin{tabularx}{\textwidth}{>{\centering\arraybackslash}m{0.1\textwidth}| >{\centering\arraybackslash}m{0.15\textwidth} |>{\centering\arraybackslash}m{0.20\textwidth} |>{\centering\arraybackslash}m{0.15\textwidth} |>{\centering\arraybackslash}m{0.233\textwidth}}
\hline
\textbf{Entity} & \textbf{Protected Object} & \textbf{Recognition Object} & \textbf{Limitation} & \textbf{Basic Requirements}\\
\hline
Data Provider & Data  & Data Source Identity & Text Quality  & Unforgeability, Robustness, Transparency, Fidelity, Radioactivity, Multi-capacity Payload\\
\hline
Technology Provider  & Model  & Model Ownership Identity  & Text Quality, Model Performance &Unforgeability, Robustness, Transparency, Fidelity, Radioactivity, Multi-capacity Payload \\
\hline
LLM User & User Privacy & User Identity, Personal Privacy  & User Capability & Unforgeability, Robustness, Watermark API\\
\hline
PRTTP &Community Ecosystem Security& AI Content Identification on Internet & Watermark Credibility & Unforgeability, Robustness, Success Rate\\
\hline
\end{tabularx}
\end{table}

\section{LLM Watermarking Technology}\label{sec3}
\subsection{Overview}
\begin{figure}[t]
    \centering
    \includegraphics[width=\linewidth]{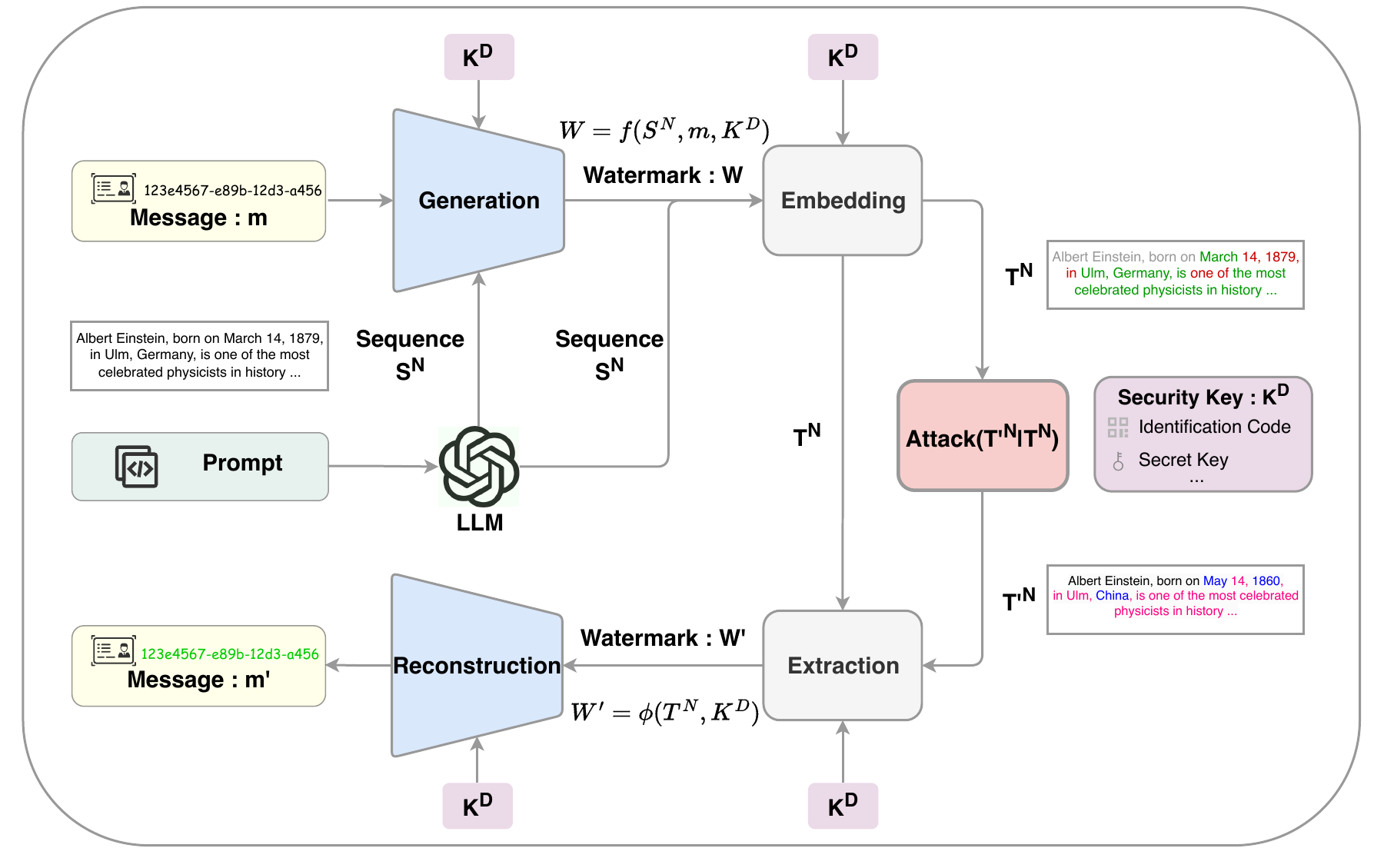}
    \caption{The watermarking technology framework in LLMs. The watermark message $m$ is used to identify the specific LLM. The security key $K^D$ represents the privacy identity tag used to generate and reconstruct the watermark. The watermark attack channels are designed to simulate attacks such as semantic substitutions and sequence changes that watermarked texts encounter during transmission. }
    \label{fig:framework}
\end{figure}

In this section, we formally define watermarking in LLMs and explore its application in securely and covertly transmitting information. Watermark algorithms for LLMs involve the processes of generation, embedding, extraction, and reconstruction, as shown in Fig.~\ref{fig:framework}. To help the readers better understand, we have listed the main symbols used in the article in the notation table~\ref{tab:notation_table}.

\begin{table}[h]
  \centering
  \begin{tabular}{c|c}
    \hline
    \textbf{Symbol} & \textbf{Description} \\ \hline
    $S^N$ & A text sequence with length N \\ 
    $T^N$ & A watermarked text sequence \\ 
    $m$ & A plain message that needs to be hidden in the original data. \\ 
    $W$ & A encrypted watermark message\\ 
    $K^D$ & An identity information key\\ \hline
  \end{tabular}
  \caption{Notation Table}
  \label{tab:notation_table}
\end{table}

We denote that the watermark message $m$, the N-vectors text sequence $S^N = (s_1, s_2, \ldots, s_N)$, the N-vectors watermarked text sequence $T^N = (t_1, t_2, \ldots, t_N)$, the D-elements watermark security parameter $K^D = (k_1, k_2, \ldots, k_D)$,  and the watermark $W$ take their values in message space $\mathcal{M}$, original sequence space $\mathcal{S}$, watermarked sequence space $\mathcal{T}$, watermark security parameter space $\mathcal{K}$, and watermark space $\mathcal{W}$ respectively. We require that $\mathcal{S}$ and $\mathcal{T}$ are isomorphic, indicating that there is a one-to-one correspondence between their elements while preserving the structure of the spaces. Moreover, $\mathcal{S}$ and $\mathcal{T}$ must obey identical distributions to ensure that the watermarking process does not alter the statistical properties of the original sequences. This is crucial for the stealth and efficacy of the watermark. Additionally, the watermark space $\mathcal{W}$ should be compatible with the original sequence space $\mathcal{S}$, where compatibility refers to the ability of the watermark signal $W$ to be embedded into the original signal $S^N$ without introducing detectable statistical differences.

A watermark $W$ is produced by the watermark generation module, followed by the generation of the watermarked text sequence $T^N$ via watermark embedding process. During text dissemination, not only are attackers likely to be present, but the message itself may also be subjected to cutting, substitution, rewriting, and reordering among other operations. An attack channel $Attack(T'^N\mid T^N)$ may apply some of the above operations to process the sequence $T^N$ to the corrupted sequence $T'^N$. The extractor, utilizing the watermark security parameter and the text sequence $T'^N$ under examination, retrieves the watermark $W'$. Subsequently, the reconstruction decodes $W'$ to calculate the estimated value $m'$ of the message initially transmitted. 


The watermarking system of LLMs can be analyzed by defining the watermark message $m$, the statistical model for the sequence $S^N$ output by the  LLM according to the prompt, and the watermark security parameter $K^D$. This includes the distortion function, constraints on the acceptable distortion levels for both the watermark embedding and the watermark attacker, and the information available to each party. The goal of the watermarking algorithm is to seek the maximum reliable transmission rate of $m$ over any possible watermarking strategy and any attack that satisfies the specified constraints. Consequently, the entire watermarking process can be described using principles of information theory. To better understand the watermarking framework, we first explain the key parameters in the diagram:

\begin{itemize}
  \item \textbf{Sequence  $S^N$:} Assume the input prompt is denoted as $Prompt$, and the sequence $S^N \in \mathcal{S}$ is composed of elements $s_1, s_2, \ldots, s_N$, where $N$ is the length of the sequence. The process by which the LLM generates a sequence can be represented as
\begin{equation}
P(S^N|Prompt) = \prod_{i=1}^{N} P(s_i \mid s_1, s_2, \ldots, s_{i-1}, Prompt).
\end{equation}

In this equation, $P(S^N \mid Prompt)$ is the probability of generating the sequence $S^N$ given the input $Prompt$. $P(s_i \mid s_1,s_2,\ldots,s_{i-1}, Prompt)$ is the conditional probability of generating the next element $s_i$ given the input $Prompt$ and the first $i-1$ elements of the sequence.

The LLM considers the sequence generated so far, $s_1, s_2, \ldots, s_{i-1}$, along with the input prompt, and then predicts the probability distribution for the next word $s_i$. Once the model predicts the probability distribution for $s_i$, it selects the next word based on this distribution, which could either be the word with the highest probability or a word sampled randomly according to the distribution. This process is repeated until the entire sequence $S^N$ is generated or a certain termination condition is met. 

\item \textbf{Watermark Message $m$:}  $m\in \mathcal{M}$ represents the message that needs to be encoded. Depending on the amount of information encoded, watermarks can be categorized into one-bit watermarks and multi-bit watermarks. 
The one-bit watermarking technique is both mature and stable; however, it is limited to encoding a single bit of information—specifically, indicating whether the text was generated by a particular LLM.
One-bit watermarks cannot meet the growing demand for customized information in LLM applications. For example, embedding model and version information in the watermark can effectively track the source of text among multiple LLMs. In contrast, multi-bit watermarks allow for carrying more customizable information. However, designing a practical multi-bit watermark method is a challenging task because multi-bit watermarks are more complex than one-bit watermarks. Consequently, embedding a multi-bit watermark can have a greater impact on the text quality compared to embedding a one-bit watermark. After the watermark generation module encodes $m$, it must be reliably transmitted to the watermark extraction module during message transmission to ensure the success rate of watermark detection. Here, $m$ is independent of $(S^N, K^D)$.

\item \textbf{Security Key $K^D$:}
$K^D \in \mathcal{K}$ represents the identity tag used to provide identity information to a text sequence. Introducing $K^D$ serves two primary purposes. Firstly, it is crucial to identify LLMs that utilize the same watermarking algorithms. This identity tag can take forms such as a secret key, providing the generator with information about the identity of the LLM that generated the text sequence $S^N$. Secondly, $K^D$ can be introduced at various stages, offering more flexibility in the identity verification process. Additionally, $K^D$ provides a known source of randomness during the extraction phase, allowing for the use of randomized codes, a standard technique to enhance transmission performance in communications.

The dependency between the original sequence $S^N$ and the identity tag $K^D$ can be quantified by the joint distribution $P(S^N,K^D)$. In public watermarks (blind watermarks), $S^N$ and $K^D$ are independent, meaning the identity tag is completely unrelated to the original text sequence. This independence implies that identity verification does not require the original text, facilitating public verification. Conversely, for private watermarks, if there is a dependency between $S^N$ and $K^D$, such as $S^N$ being a function of $K^D$, validating the identity tag requires access to the original text sequence or the original encoding parameters. 
\end{itemize}




\begin{figure}[!ht]
    \centering
    \includegraphics[width=\linewidth]{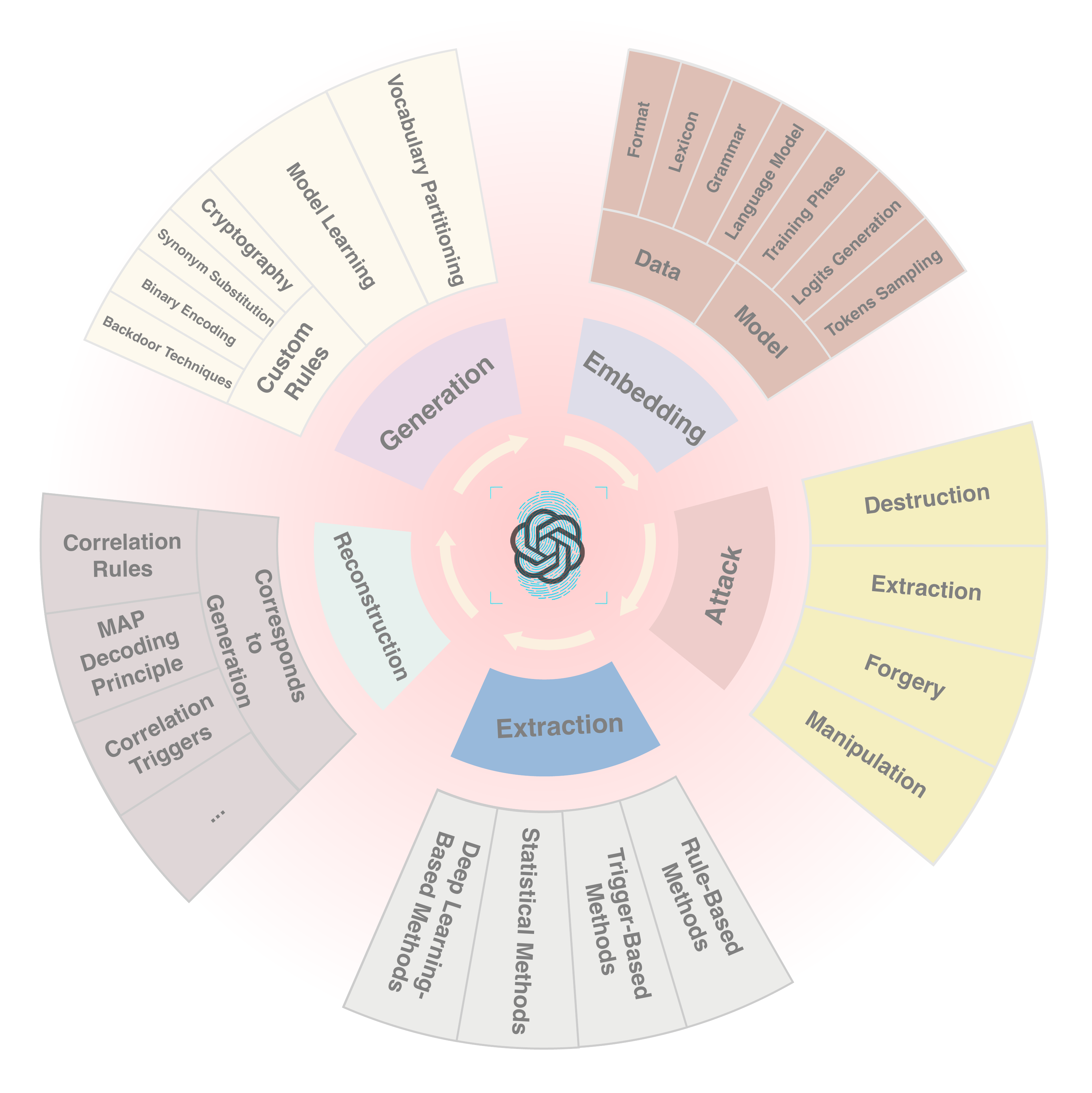}
    \caption{The overview of watermark algorithms in LLMs.}
    \label{fig:overview}
\end{figure}

\subsection{Problem Definition} \label{pd}
Different LLM watermarking algorithms have distinct parameters and settings at each stage, playing various roles throughout the watermarking process.
The existing LLM watermark algorithms are summarized in Fig.~\ref{fig:overview} categorized by the different phases of the watermarking process.

\textbf{\romannumeral1) Watermark generation: }The watermarking algorithm encodes the text \( S^N \) generated by an LLM, the watermark security identity \( K^D \), and the watermark message \( m \) through the function $f$. Initially, the watermark information $m$ must be converted into a feature suitable for embedding, generating the corresponding watermark signal $W$. The method of generation simultaneously affects the watermark's information capacity, transparency, robustness, and other indicators. In LLM watermarking, the entire generation stage aims to disperse the watermark information $m$ into the feature space of the sequence, mapping the LLM output text sequence $S^N$, the watermark message $m$, and the watermark security parameter $K^D$ to the watermark signal $W$, which can be embedded into the original sequence space $\mathcal{S}$ and satisfy certain constraints. The generation process is denoted as
\begin{equation}\label{formula:W}
W = f(S^N, m, K^D)
\end{equation}
\begin{equation}
f: \mathcal{S} \times \mathcal{M} \times \mathcal{K} \to \mathcal{W}.
\end{equation}

The function $f$ is the mapping of the sequence $S^N$ generated by the LLM, the watermark message $m$, and the watermark security parameter $K^D$ to the watermark signal $W$.

We define two key concepts: \textit{Attack Robustness} and \textit{Security Robustness}. From an information-theoretic perspective, these concepts for the watermark signal $W$ can be characterized by mutual information $I$. \textit{Attack Robustness} represents the ability to withstand attacks against the watermark. A larger $I(S^N; W)$ indicates a stronger dependency between the generated watermark $W$ and the original sequence, signifying a higher capability to resist watermark attacks. Conversely, \textit{Security Robustness} refers to the capacity to prevent the deduction of the watermark message from the watermarked sequence $T^N$. A smaller $I(m; W)$ signifies a weaker dependency between the generated watermark $W$ and the watermark message $m$, thereby inhibiting the inference of the watermark message $m$ from $W$.

To simultaneously consider \textit{Attack Robustness} and \textit{Security Robustness}, the optimization goal is 
\begin{equation}\label{eq4}
\begin{aligned}
& \underset{f_N}{\text{max}} \quad I(S^N;W)-\lambda I(m;W) \\
& \text{s.t.} \quad
\begin{aligned}
& (i) \quad W = \arg\max_{f_N} I(S^N;W)-\lambda I(m;W) \\
& (ii) \quad I(m;W) \leq \epsilon,
\end{aligned}
\end{aligned}
\end{equation}
which aims to maximize the mutual information $I(S^N;W)$ between the original sequence $S^N$ and the watermark $W$, while minimizing the mutual information $I(m;W)$ between the watermark message $m$ and the watermark $W$. Ideally, $I(m; W)$ should be zero, indicating that $m$ and $W$ are completely uncorrelated, thereby achieving complete transparency of the watermark. The parameter $\lambda$ is a positive trade-off coefficient that adjusts the balance between these two objectives. Furthermore, $\epsilon$, a very small positive number, quantifies the security robustness of the watermark.

\noindent\textbf{\romannumeral2) Watermark Embedding:} After obtaining the watermark signal $W$ through the generation phase, it is necessary to embed $W$ into the watermark carrier (i.e., the original text sequence $S^N$) to produce the watermarked text sequence $T^N$. We define the operation of embedding the watermark as the function $Emb$:
\begin{equation}
T^N = Emb(S^N,W).
\end{equation}
The $Emb$ function may employ simple techniques such as addition, concatenation, or more complex watermark embedding operations. These could include embedding the watermark $W$ from various perspectives, such as format, vocabulary, and syntax at the data level or by manipulating the training and inference processes of LLMs at the model level. Integrating Formula ~\ref{formula:W}, the embedding operation can be expressed as $$ T^N = Emb(S^N, f(S^N, m, K^D)). $$

From the perspective of watermark \textbf{\textit{Text Quality}}, the mutual information $I(S^N; T^N)$ between the embedded text sequence $T^N$ and the initial sequence $S^N$ should be maximized. Considering the watermark \textbf{\textit{Transparency}}, the correlation between the generated watermark signal $W$ and the embedded text sequence should be as small as possible, i.e., $I(W; T^N)$ should be minimized. Therefore, the optimization objective can be defined as
\begin{equation}\label{eq6}
\begin{aligned}
& \underset{Emb}{\text{max}}\quad I(S^N;T^N)-\theta I(W;T^N) \\
& \text{s.t.} \quad
\begin{aligned}
& E\{d_N(S^N, {T}^N)\} = \sum_{S^N \in \mathcal{S}} \sum_{K^D \in \mathcal{K}} \sum_{m \in \mathcal{M}} \frac{1}{|\mathcal{M}|}P(S^N, K^D) \\
&\times d_N(S^N, T^N) \leq D_{emb}.
\end{aligned}
\end{aligned}
\end{equation}


The watermark \textbf{\textit{Transparency}} is also subject to an average distortion constraint $D_{emb}$. The definition of the distortion constraint involves an average over the distribution $p(S^N, K^D)$ and a uniform distribution over the messages. A non-negative bounded distortion function $d_{emb}(x_i, y_j) = \begin{cases} 
0, & x_i = y_j \\
a, & a > 0, x_i \neq y_j 
\end{cases}$
exists between elements of the sets $\mathcal{S}$ and $\mathcal{T}$. This average distortion $D_{emb}$ is the value of the average distortion 
\begin{equation}
\bar{D}_{emb} = \sum_{S^N \in \mathcal{S}} \sum_{T^N \in \mathcal{T}} p(T^N) p(S^N \mid T^N) d_N(S^N, T^N).
\end{equation}
 The distortion function is extended to $N$-dimension-vectors by $d_N(S^N, {T}^N) = \frac{1}{N} \sum_{i=1}^N d_{emb}(s_i, t_i).$
This constraint further limits the degree of distortion in the text sequence with the embedded watermark ensuring the transparency of the watermark.

Watermark embedding techniques can be classified into two main categories: data-centric watermark embedding and model-centric watermark embedding. Each methodology has distinct features and application contexts, collectively laying the technological foundation for the protection of intellectual property, model security, and the authentication of data and models.

\noindent\textbf{\romannumeral3) Watermark Extraction: }The function $\phi :\mathcal{T} \times \mathcal{K} \to \mathcal{W}'$ is the extractor mapping, which takes the watermarked text $T^N$ and the watermark security parameters $K^D$, and maps them to the extracted watermark signal $W'$:
\begin{equation}
W' = \phi(T^N,K^D).
\end{equation}

At this stage, we revisit the watermark embedding process of text length $N$, constrained by distortion $D_{emb}$, which can be defined as a triplet $(\mathcal{M}, f, \phi)$ where: $\mathcal{M}$ is the watermark message space.

\noindent\textbf{\romannumeral4) Watermark Reconstruction: }After extracting the watermark signal $W'$, it is necessary to decode the watermark message $m'$ from $W'$. To approach the channel capacity with a reliable transmission rate of the watermark message, a jointly optimal decoding rule, designed corresponding to the generation phase, should be adopted to compute the estimated value of the original watermark message $m'$.

If the Maximum A Posteriori (MAP) decoding principle is adopted to minimize the error probability:

\begin{equation}\label{eq8}
m' = \arg \max_{m \in \mathcal{M}} p(m\mid W', K^D).
\end{equation}

Other decoding rules can also be adopted, such as correlation rules, normalized correlation rules, or trigger-based rules.

\noindent\textbf{\romannumeral5) Watermark Attacks:}
With the advancement of watermarking technologies, attack methods targeting watermarks have also evolved. These attacks aim to undermine the effectiveness of watermarks and, in some cases, completely remove them, posing a threat to content security and copyright maintenance. Notably, these methods of attack not only represent potential threats but are also utilized to assess the robustness of watermarking technologies. This, in turn, helps developers improve and fortify watermark algorithms. 

\textbf{Adversary’s capabilities.}  If watermark attacks occur throughout the watermarking algorithm cycle, we consider an adversary with black-box input-output access to the language model. 
In public watermark mode, the adversary is aware of all the details of the public algorithm. In private watermark mode, the adversary knows the watermark implementation but lacks knowledge of the security key $K^D$ and the encryption component of the watermark generation algorithm.

This adversary has the capacity to modify the sequence $T^N$ within a distortion constraint. Given the distortion function between elements of the sequence spaces $\mathcal{S}$ and $\mathcal{S}'$, denoted as $d_{atk}(\cdot, \cdot)$, subject to the distortion constraint $D_{atk}$. The attack channel  $\text{Attack}(T'^N|T^N)$ is defined as a sequence of conditional probability mass functions (p.m.f.) from space $\mathcal{S}$ to $\mathcal{S}'$, where the distortion is evaluated relative to the original LLM sequence $S^N$ rather than the watermarked sequence $T^N$:
\begin{equation}\label{eq9}
E\{d_{atk}(T^N, T'^N)\} = \sum_{m,S^N,T^N,K^D,T'^N} d_{atk}(S^N, T'^N)P(T'^N\mid T^N)P(S^N, K^D) \leq D_{atk}.
\end{equation}

\textbf{Adversary’s objective.}  The primary objective of the adversary is to render the watermark extraction algorithm ineffective. Specifically, the adversary aims to produce a $T'^N$ such that $\phi(T'^N, K^D) \neq W$, while ensuring that  $T'^N$ remains a minor modification of the LLM-generated watermark sequence $T^N$.

\subsection{Watermark Generation}\label{sec_generation}

\subsubsection{Vocabulary-partitioning-based Methods}
Watermarks generated through vocabulary partitioning usually utilize a pseudo-random function (implemented as a hash function) to generate random seeds. These seeds are used to divide the vocabulary into distinct lists, ensuring that a subset of tokens from a particular list is output more frequently during token generation. The watermark is generated by biasing the selection of tokens towards specific lists.

\begin{figure}[H]
    \centering
    \includegraphics[width=\linewidth]{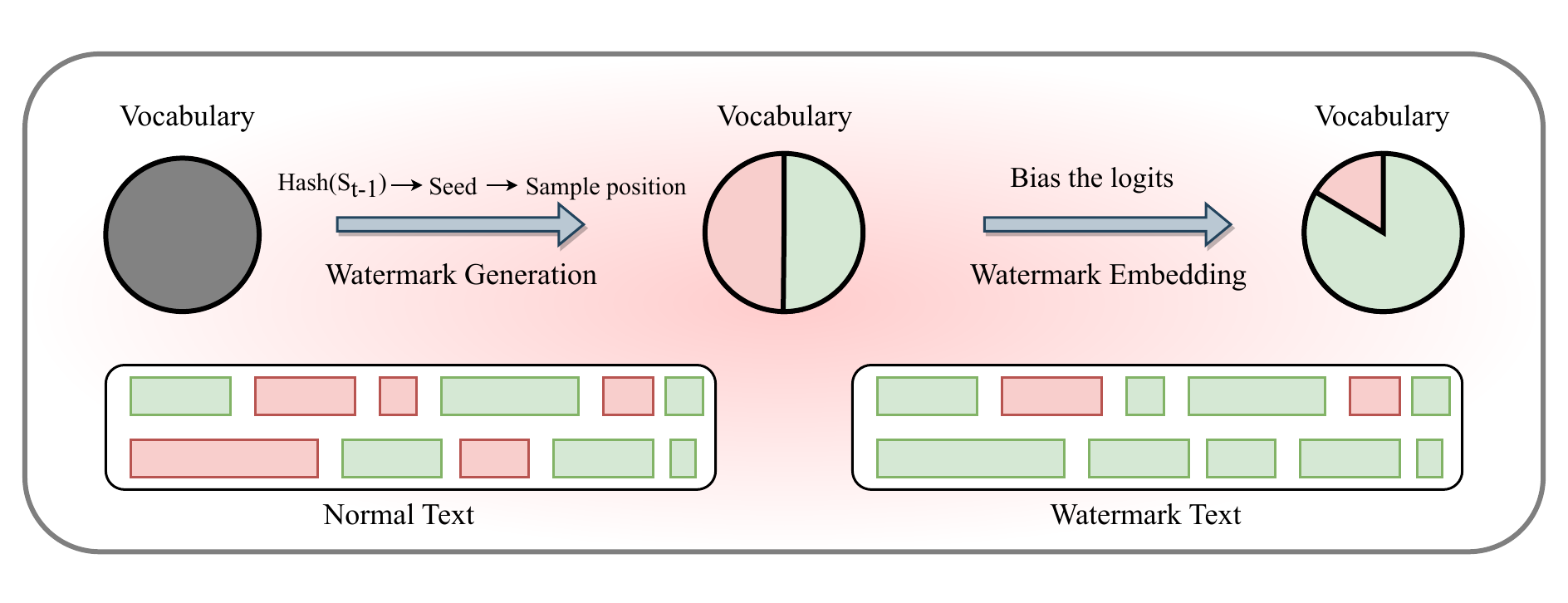}
    \caption{Watermark generation through vocabulary partitioning. Utilizing a hash function, the previous token is used as input to compute a random seed, which divides the vocabulary into green and red lists. The LLM-generated token bias is applied by adding a bias term to the token log probabilities, favoring the green list.}
    \label{fig:enter-label}
\end{figure}

Kirchenbauer et al.~\cite{kirchenbauerWatermarkLargeLanguage2023} introduced the first LLM watermarking technique that generates watermarks through vocabulary partitioning, as shown in Fig.~\ref{fig:enter-label}. This method employs a hash function that uses the preceding token as input to compute a random seed for partitioning the vocabulary into green and red lists. The LLM-generated tokens are biased towards the green list by adding a bias term to the logits. During detection, the extraction process calculates the ratio of green tokens with the z-metric to determine the presence of the watermark.

Following Kirchenbauer et al.~\cite{kirchenbauerWatermarkLargeLanguage2023}, the idea of watermarking generation by vocabulary partitioning has been widely explored by many researchers~\cite{kirchenbauerReliabilityWatermarksLarge2023,takezawaNecessarySufficientWatermark2023,huUnbiasedWatermarkLarge2023,zhaoProvableRobustWatermarking,wuDiPmarkStealthyEfficient2023,fuWatermarkingConditionalText2024,leeWhoWroteThis2024,renRobustSemanticsbasedWatermark2024}. These methods have introduced more refined methods for partitioning the red-green lists, resulting in a greater diversity of partitioning techniques. For example, Takezawa et al.~\cite{takezawaNecessarySufficientWatermark2023} introduced tighter constraints for partitioning red-green tables,enhancing the concealment of the watermark and improving the quality and naturalness of the generated text. Building upon previous work, Kirchenbauer further elaborated on the robustness of these watermarking generation methods against paraphrasing attacks~\cite{kirchenbauerReliabilityWatermarksLarge2023}. The effectiveness of the detection method in this study was evaluated by comparing them with detectors designed to identify AI-generated text. 

While these watermarks can be integrated with various detection techniques, their distribution does not meet the unbiased criterion. To address this, Hu et al.~\cite{huUnbiasedWatermarkLarge2023} introduce an unbiased watermark by adjusting the token generation probability distribution through the watermark code space. Some studies focus on enhancing the robustness of such watermarks against attacks, aiming to mitigate vulnerabilities arising from reliance on lexical distribution. For instance, Li et al.~\cite{liDoubleIWatermarkProtecting2024} employed a novel reweighting strategy combined with a context-based hash function to assign a unique i.i.d. ciphers to each generated token. This encoding method ensures the preservation of the original token distribution during the watermarking process, making the watermarked text indistinguishable from unwatermarked text in terms of distribution. Other studies~\cite{fuWatermarkingConditionalText2024,renRobustSemanticsbasedWatermark2024} have considered semantic similarity when partitioning the vocabulary, such that the semantic value may remain unchanged even in the face of watermark attacks, thereby achieving robustness against paraphrasing attack. 

Watermark generation methods based on vocabulary partitioning rely on hashing tokens from the previous moment, leading to inefficiency during the extraction phase, as it necessitates iterative computation over all tokens.
To address this, Zhao et al.~\cite{zhaoProvableRobustWatermarking} simplified the watermarking generation method proposed by Kirchenbauer et al. by employing a fixed increase in logits watermark strength $\epsilon$, making the vocabulary partitioning independent of previously generated tokens and solely reliant on a global key. Fernandez et al.~\cite{fernandezThreeBricksConsolidate2023} proposed a method to enhance efficiency by cyclically shifting an initial message to generate secret vectors for each message, thereby easily converting one-bit watermarks into multi-bit watermarks and allowing for parallel processing. Additionally, Liu et al.~\cite{liuUnforgeablePubliclyVerifiable2024} employed a watermark generation network to partition the vocabulary instead of using hash functions, which has also been proven effective.

Some studies have extended the method of partitioning the vocabulary into multi-bit watermarks to convey more information through the watermark. However, these methods also face higher computational complexity and increased demands for watermark information density.
To develop a more effective multi-bit watermark techniques, Wang et al.~\cite{wangCodableWatermarkingInjecting2023b} proposed the Balance-Marking method, which uses a proxy language model (proxy-LM) to ensure that the available and unavailable vocabulary for generating watermarks have approximately equivalent probabilities. 
Similar to the work of Lee et al.~\cite{leeWhoWroteThis2024}, this method can also circumvent low-entropy parts of the text to effectively improve text quality.

However, this watermarking approach essentially divides the vocabulary into multiple sets of red-green tables, with each set corresponding to one bit of watermark message. Such methods requires iterative computation during the logits generation process for each token, resulting in extremely high computational complexity. To mitigate this, besides the cyclic shift method by Fernandez et al.~\cite{fernandezThreeBricksConsolidate2023}, Yoo et al.~\cite{yooAdvancingIdentificationMultibit2024} independently encode each message bit position, transforming the division of the vocabulary from red-green lists into colored lists, effectively encoding multiple states for every token. Allocating tokens to different parts of the message allows for embedding longer messages without increasing generation latency. Compared to methods that directly generate watermarks using the watermark message $m$, Qu et al.~\cite{quProvablyRobustMultibit2024} have enhanced the robustness of the watermark by introducing error-correcting codes (ECC) to the watermark information before dividing the vocabulary. 

The key limitation of the existing multi-bit watermark approaches~\cite{fernandezThreeBricksConsolidate2023,wangCodableWatermarkingInjecting2023b} is that the computational cost of their extraction functions grows exponentially with the length of the watermark message bits, and they cannot accurately or effectively extract all watermark bits. 


 The methods of vocabulary partitioning involve mapping the text sequence $S^N$ to various distributions of vocabulary that can be analyzed for their $Attack$ $Robustness$ and $Security$ $Robustness$ using Formula~\ref{eq4}. All these methods incorporate semantics, global secret keys, and additional information to further solidify the dependency between the generated watermark $W$ and the original sequence $S^N$, which increases $I(S^N; W)$. The original KGW~\cite{kirchenbauerWatermarkLargeLanguage2023} partitions the vocabulary using the hash value of previous tokens. The hash function ensures the mutual information $I(W; m) \to 0$ between the watermark $W$ and the watermark message $m$, making this method a high level of $Security$ $Robustness$. Denote that the mutual information $I(m; W) = H(m) - H(m\mid W)$, where $H(m)$ is the entropy of the message $m$ and $H(m \mid W)$ is the conditional entropy of $m$ given the watermark $W$. Since the computation process of hash functions is unidirectional and irreversible, the conditional entropy $H(m\mid W)$ encloses to $H(m)$ when the output of the hash function $W$ is known. Therefore, the mutual information $I(m; W) \to 0$ is minimized to prevent an attacker from back-propagating the message $m$ through the watermarked signal $W$. 
 
 The work~\cite{kirchenbauerReliabilityWatermarksLarge2023} uses a context-robust Min-Left Hash to strengthen the connection between $W$ and the original sequence $S^N$, thereby increasing $I(S^N; W)$. Zhao et al.~\cite{zhaoProvableRobustWatermarking} no longer use a hash function for vocabulary partitioning but instead base it on text edit distance to partition a fixed vocabulary. Although the fixed vocabulary has a higher $I(W; m)$ than the hash-partitioned vocabulary, its $Security$ $Robustness$ is reduced. However, by increasing the edit distance between vocabularies, attackers need multiple attempts to bridge the text distance and invalidate the watermark, thus enhancing $Attack$ $Robustness$. 
 
 Additionally, for low-entropy texts mentioned in paper~\cite{kirchenbauerWatermarkLargeLanguage2023}, which are difficult to watermark by modifying logits, this can also be analyzed using Formula~\ref{eq4}. When the original sequence $S^N$ has low entropy, the first term of the optimization goal, $I(S^N; W)$, is low, which negatively impacts watermark generation and reduces its $Attack$ $Robustness$. Consequently, some studies~\cite{leeWhoWroteThis2024, quProvablyRobustMultibit2024} propose bypassing low-entropy texts and only watermark high-entropy texts to ensure the watermark's $Attack$ $Robustness$. Furthermore, semantic-based watermarking methods determine vocabulary partitioning by incorporating contextual relationships and semantic information. The introduction of semantic information enhances the correlation between the generated watermark $W$ and the original sequence $S^N$. The increase in $I(S^N; W)$ enhances the robustness of such methods against watermark attacks. For instance, Fu et al.~\cite{fuWatermarkingConditionalText2024} selected semantically related vocabulary to add to the watermark vocabulary. Ren et al.'s SemaMark~\cite{renRobustSemanticsbasedWatermark2024} mitigated semantic sensitivity by discretizing the continuous word embedding space appropriately, ensuring that discrete semantic values remain unchanged even in the face of watermark text editing attacks.

\subsubsection{Model-learning-based Methods}
Model-based learning methods employ deep learning techniques, such as GPT~\cite{radfordImprovingLanguageUnderstanding2018}, BERT~\cite{devlinBERTPretrainingDeep2019}, to generate watermarks. 
 These methods leverage the learning and generative capabilities of deep learning models, using a trained watermark generation model to directly produce watermarks $W$ or embedded representations of watermarked sequences. 
 
 In contrast to other methods, this technique create the watermark that is intricately embedded into the content, enhancing security and robustness against tampering. Correspondingly, watermark extraction is typically performed using a dedicated decoder or a watermark detection network. This dual-model framework ensures that the embedded watermarks can be accurately retrieved, even when the content has undergone modifications or compression. By maintaining the watermark generation model as proprietary while making the watermark detection model publicly accessible, a publicly verifiable watermarking scheme can be effortlessly implemented. Conversely, by keeping both the watermark generation and detection models confidential, the watermarking method can be transformed into a private watermarking system. 
 
 This strategic dichotomy allows for flexibility in controlling the accessibility and verification of watermarks, catering to different security and privacy requirements. Publicly verifiable watermarks facilitate widespread verification, enhancing transparency and trust in digital content authenticity. In contrast, private watermarking schemes offer enhanced security since the ability to generate and detect watermarks is restricted to authorized entities. This restriction safeguards proprietary or sensitive information from unauthorized detection and manipulation.

Kuditipudi et al.~\cite{kuditipudiRobustDistortionfreeWatermarks2023} employed a decoder that deterministically maps a sequence of random numbers, encoded by a watermark key, to samples in a language model. This is achieved by converting a sequence of uniform random variables and permutations into tokens using inverse transform sampling. Considering that many existing watermarking algorithms are designed at the token level, Hou et al.~\cite{houSemStampSemanticWatermark2023} utilized a sentence encoder trained through contrastive learning (such as Sentence-BERT~\cite{reimersSentenceBERTSentenceEmbeddings2019}) to capture textual semantic similarities. They partitioned the semantic space of sentences and employed sentence-level rejection sampling to ensure that sentences fall within watermarked partitions of this space. This approach to semantic watermarking at the sentence level shows strong robustness against paraphrasing attacks. Munyer et al.~\cite{munyerDeepTextMarkDeepLearningDriven2024} utilized Word2Vec~\cite{church2017word2vec} and Sentence Encoding~\cite{cer2018universal} to engender a roster of replacement words, which we consider as the generated watermark.

The mainstream method of embedding watermarks is to add extra watermark logits on top of the logits generated by the LLM. Some intriguing research~\cite{liuUnforgeablePubliclyVerifiable2024,liuSemanticInvariantRobust2024}, inspired by the red-green list method~\cite{kirchenbauerWatermarkLargeLanguage2023}, moves away from guiding logit modifications by partitioning the vocabulary. Instead, these studies directly generate watermark logits through a watermark generation model. Most watermarking methods cannot simultaneously possess $Attack$  $Robustness$ against watermark attacks and $Security$ $Robustness$ to prevent inferring the watermark from the watermarked sequence $T^N$, necessitating a trade-off. The research by Liu et al.~\cite{liuSemanticInvariantRobust2024} makes the generated watermark $W$ no longer determined by previous tokens and vocabularies, thereby enhancing both $Attack$ and $Security$ $Robustness$. Gu et al.~\cite{guLearnabilityWatermarksLanguage2024} took an alternative approach by training a student model to learn the token distribution of watermarked text, imitating the behavior of existing watermarking algorithms through model distillation. However, this approach involves model distillation and suffers from high computational complexity.

Several pioneer works~\cite{abdelnabiAdversarialWatermarkingTransformer2021,yooRobustMultibitNatural2023,zhangREMARKLLMRobustEfficient2023,wangCodableWatermarkingInjecting2023b} have explored designing multi-bit watermark schemes for LLMs by leveraging the model's learning capabilities. Abdelnabi and Fritz~\cite{abdelnabiAdversarialWatermarkingTransformer2021} proposed an encode-decoder transformer architecture, AWT, which learns to extract the message from the decoded watermarked text. To maintain the quality of the watermarked text, they utilize signals from sentence transformers and language models, relying entirely on a neural network for message embedding and extraction. This approach has proven effective because neural networks have been successfully used for natural language watermarking, demonstrating their capability to handle complex language patterns.

Drawing inspiration from a well-known proposition in classical image watermarking work~\cite{coxSecureSpreadSpectrum1997}, Yoo et al.~\cite{yooRobustMultibitNatural2023} generated watermark positions by identifying invariant features of semantics and syntax in the text through a pre-trained infill model and create watermarks by replacing words at these watermark positions through masking. Due to the use of a semantically robust filling model, their method significantly surpasses AWT in resilience to watermark attacks and exceeds the fixed upper limit on the number of watermark bits imposed by the ContextLS method proposed by Yang et al.~\cite{yangTracingTextProvenance2022a}. Extending AWT, Zhang et al.~\cite{zhangREMARKLLMRobustEfficient2023} utilized pre-trained language models in a modular fashion to revamp the end-to-end watermarking scheme. They introduced a reparameterization module to transform the dense distributions from the message encoding to the sparse distribution of the watermarked textual tokens, achieving double the watermark information capacity of AWT while maintaining semantic integrity.

Wang et al.~ \cite{wangCodableWatermarkingInjecting2023b} systematically studied the codable watermark system (CTWL) for multi-bit watermark information, considering Yoo et al.~\cite{yooRobustMultibitNatural2023}'s watermarking method as post-process after LLM generation, which does not integrate well with the generative capabilities of LLMs. They proposed using a proxy language model (proxy-LM) to assist in encoding watermark information during the LLM generation process, followed by vocabulary division.

 As indicated by Formula~\ref{eq4}, these methods utilize the learning and semantic capabilities of the model to generate the watermark $W$. For instance, Liu et al.~\cite{liuSemanticInvariantRobust2024} utilized a trained watermark model to generate watermark logits based on the semantic embeddings of tokens preceding the current token, which maximizes the first term $I(S^N; W)$ as much as possible. It is understood that the information flow in neural networks tends to decrease mutual information during forward propagation~\cite{belghaziMutualInformationNeural2018}. This decrease is due to the nonlinearity of forward propagation, the many-to-one mapping relationship, and the suppressive effect of activation functions on information flow in neural networks~\cite{davisNIFFrameworkQuantifying2019}. Inferring the input from the network's output is very difficult, thereby making the conditional entropy $H(m|W)$ encloses to $H(m)$, which ensures that $I(m, W) \to 0$. This also ensures the $Security$ $Robustness$ of such watermarking methods.


  


\subsubsection{Custom-rules-based Methods}
The methods proposed in this section involve generating watermarks by applying specific rules. The core principle is to design a set of rules or algorithms that modify or mark text data and models, thereby generating watermarks.

\textbf{Backdoor Techniques:}
These methods involve poisoning the training data of LLMs by adding specific triggers to text sequences, thus enabling LLMs to learn the characteristics of these triggers. The presence of watermarks is detected by observing the output of LLMs when given input samples containing embedded triggers. 

For instance, Liu et al.~\cite{liuWatermarkingClassificationDataset2023} implanted backdoor triggers into a small subset of the target LLM's training data and tampered with the labels of this subset. The presence of watermarks is assessed by verifying the output of the trigger set through black-box access to the target model. Tang et al.~\cite{tangDidYouTrain2023}improved the backdoor poisoning method without altering the original labels of the watermark samples. Instead, they guided the model to memorize the preset backdoor function by disabling original features on watermark samples through imperceptible perturbations. This approach increases the transparency of the watermark and, by protecting the trigger set used for watermark verification, creates a traceable private watermarking technology. These methods are primarily data-driven, reliant on data, and require participation in the training process of LLMs, with a limited capacity to carry one-bit watermark information.

\textbf{Cryptography:}
Watermark generation based on cryptography aims to enhance the security and stealthiness of watermarks through cryptographic techniques. These methods primarily include the use of digital signature technology for watermarks~\cite{fairozePubliclyDetectableWatermarking2023} and cryptographically inspired undetectable watermarks~\cite{christUndetectableWatermarksLanguage2023}, both of which rely on cryptographic principles to protect the watermark from unauthorized access and tampering. Christ et al.~\cite{christUndetectableWatermarksLanguage2023} quantified the randomness used in the generation of a specific output by utilizing pseudo-random values generated by an encrypted pseudo-random function (PRF) to determine watermark embedding locations. They analyzed the undetectability and integrity of the watermark using empirical entropy theory. Furthermore, they introduced the encryption of pseudo-random values with a secret key, which is required for the extraction and verification of the watermark. However, this method is only validated through the entropy theorem for binary channels and is limited to embedding one-bit watermark information. It is uncertain whether it can maintain sufficient empirical entropy to ensure the robustness of the watermark when expanded to multiple bits of information.

Additionally, Fairoze et al.~\cite{fairozePubliclyDetectableWatermarking2023} explored the application of digital signature technology on LLMs.This method involves encrypting the hash value of text with a private key to create a watermark signature. This signature is then embedded into tokens of additional length through rejection sampling, while the public key facilitates watermark detection. This approach does not require embedding statistical signals in the generated text, providing a viable solution for publicly detectable LLM watermarks. However, this approach, which employs asymmetric algorithms, often results in highly unstable time and computational costs during watermark generation. The running time exhibits high variance, especially when encountering low-entropy sections of sampling or missed hashes, making the time required for watermark generation occasionally unacceptable.

The main advantage of incorporating cryptographic techniques is the ability to determine whether a watermark is private or publicly detectable easily. One primary benefit of public watermarks is that the extraction and reconstruction processes can be outsourced, allowing different entities to provide watermark extraction services separate from the model providers. Furthermore, public watermark schemes should support the full lifecycle operations of watermarks through API access to private LLMs.



 \textbf{Custom Synonym Substitution Rules:} Some methods~\cite{heCATERIntellectualProperty,yangTracingTextProvenance2022,liProtectingIntellectualProperty2023} ensure a close relationship between the watermark and the text's semantics and context by using custom synonym replacement rules. This approach not only maintains the transparency of the watermark but also enhances its $Attack$ $Robustness$ to text editing attacks. The fundamental premise of text editing attacks is to invalidate the watermark under certain distortion conditions. However, effectively linking semantics and context can limit the effectiveness of such attacks. For instance, He et al.~\cite{heCATERIntellectualProperty} considered two fundamental linguistic features during synonym replacement: part-of-speech (POS) and the dependency tree. Expanding to multi-bit watermarks, Yang et al.~\cite{yangTracingTextProvenance2022} proposed a context-aware synonym replacement method for generating watermarks.  Meanwhile, Li et al.~\cite{liProtectingIntellectualProperty2023} embedded a series of synonym-based token changes as watermarks in the code generated by LLMs. Li et al.~\cite{liResilientWatermarkingLLMGenerated2024} introduced a watermarking method for code text based on transformation rules such as code refactoring, reordering, and format conversion. Each transformation corresponds to a bit of the watermark message, with the presence or absence of a specified transformation determining whether the watermark value of that bit is 0 or 1.

 \textbf{Custom Generation Function: } Some custom watermark generation functions have proven effective in practice. For instance, Yang et al.~\cite{yangWatermarkingTextGenerated2023} defined a binary encoding function for black-box LLMs, which calculates a random binary code (0 or 1) for each word in the text based on the hash value of the word and its preceding word. Essentially, the function of this binary encoding is similar to that used by Kirchenbauer et al.~\cite{kirchenbauerWatermarkLargeLanguage2023}, who utilized hash functions for vocabulary partitioning. Zhao et al.~\cite{zhaoProtectingLanguageGeneration2023} defined two sets of specialized secret sinusoidal signals as watermarks. These two sets of sinusoidal signals have values ranging from [0,1] and satisfy the condition that their sum equals 1.

\subsection{Watermark Embedding}\label{sec_embedding}
After generating the watermark $W$, it must be embedded into the sequence carrier $S^N$. Depending on the direct object of operation during the embedding process, the embedding can be divided into two types: data-level embedding and model-level embedding.

\subsubsection{Data-level Embedding Methods}
Data-level embedding, also known as post-processing methods, involves inserting watermarks by directly modifying, learning from, or augmenting the content itself. The primary advantage of these methods is their ability to embed identifying markers $W$ within data in a relatively concealed manner without requiring modifications to the model's architecture or functionality. Data embedding methodologies exhibit a broad spectrum, including format adjustments, lexical changes, grammatical shifts, and the exploitation of language models, thus offering a variety of approaches to suit different data types and application contexts. Focused on the textual level, these methods are not only applicable to the training data of LLMs to influence the learning trajectory but can also be directly applied to the text generated by LLMs, enabling watermark embedding at the output phase. This phase is principally segmented into four categories:

\textbf{Format Adjustment:} Format-based data embedding ingeniously utilizes text formatting and visual features for watermark embedding. Unlike methods that directly modify the text content, format-based methods embed watermarks through subtle adjustments to the appearance and structure of the text, aiming to achieve copyright protection and data tracking without compromising readability and content integrity.

Btassil et al.~\cite{brassilElectronicMarkingIdentification1995}achieved watermark embedding by adjusting the vertical and horizontal positions of text lines and words, such as through line shift coding and word shift coding. Por et al.~\cite{porUniSpaChTextbasedData2012} embedded watermarks by inserting different space characters into text spacing or using visually similar but differently coded characters. The presence of these watermarks is almost invisible to users, ensuring the natural flow and original appearance of the text content. These methods do not rely on changes to the text content, making it broadly applicable across various languages and document types without concerns about linguistic or semantic restrictions.
However, watermarks embedded in this way have poor robustness and can be invalidated by some formatting-checking tools.

\textbf{Lexical Variation:} Lexicon-based data embedding is a method that embeds watermarks by carefully selecting and replacing specific words in the text. This approach leverages the richness and diversity of language, allowing for subtle modifications without altering the original intent and content. 

Aside from the watermark generation method~\cite{heCATERIntellectualProperty,yangTracingTextProvenance2022,liProtectingIntellectualProperty2023} of Custom Synonym Substitution Rules mentioned in Section~\ref{sec_generation}, which embeds watermarks through vocabulary changes, the embedding of watermarks through lexical variation is also widely used by various watermark algorithms~\cite{keskisarkka2012automatic,abdelnabiAdversarialWatermarkingTransformer2021,heProtectingIntellectualProperty2022,munyerDeepTextMarkDeepLearningDriven2024,yangWatermarkingTextGenerated2023,yooRobustMultibitNatural2023,qiangNaturalLanguageWatermarking2023}.

Lexicon-based data embedding ensures both fluency in text reading and semantic consistency. This characteristic renders some watermark detection methods ineffective and ensures the transparency of the watermark, making it a mainstream method in many algorithms. However, this approach relies on high-quality synonym databases and advanced language models or rules for precise vocabulary selection and sentence transformation.

\textbf{Grammatical Transformation:} Grammar-based data embedding is a technique that embeds watermarks by altering the syntactic structure of text or code. This method aims to incorporate watermark information through subtle syntactic adjustments without affecting the original semantics. Its application is not limited to natural language texts but also extends to programming languages, demonstrating wide applicability and flexibility.

Chalmers~\cite{chalmersSyntacticTransformationsDistributed} inserts watermarks by transforming the syntactic structure of sentences within paragraphs. 
In the CATER method proposed by He et al.~\cite{heCATERIntellectualProperty}, the dependency tree is a type of syntactic structure that describes the directed binary grammatical relationships between words.

These embedding methods adjust the text at the grammatical level, maintaining semantic integrity and high transparency, and are commonly used for watermark embedding in code text. However, they require a deep understanding of grammar and analysis capabilities. Excessive grammatical changes can affect text readability.

\textbf{Language Model Utilization:} This embedding method further leverages the capabilities of language models. Most watermark algorithms generated through model learning primarily utilize the semantic understanding abilities of language models to create watermarks, necessitating other embedding methods. In contrast, Zhang et al.~\cite{zhangREMARKLLMRobustEfficient2023} directly trained a message encoding module that takes watermark messages as input and generates watermarked text based on the learning capabilities of language models. This end-to-end training approach fully exploits the capabilities of language models. Since the watermark is directly generated and embedded internally by the language model, its robustness against text editing attacks depends on the robustness of the language model itself. The generation and embedding of the watermark rely on the model, and if data changes, it necessitates adjusting the model training objectives, which requires substantial computational resources.

The embedding methods at the data level involve manipulating text sequences generated by LLMs, with the fundamental principle of preserving the original sequence's semantics and readability. According to Formula~\ref{eq6}, the text quality represented by $I(S^N; T^N)$ is assured. However, detailed text analysis of vocabulary, format, grammar, etc., poses a risk of inferring the watermark signal $W$ from the watermarked text $T^N$. Therefor, that  watermark embedding method has not effectively minimized $I(W; T^N)$.

\subsubsection{Model-level Embedding Methods}
The model-level embedding approach differs from data-level methods by embedding watermarks directly within the application cycle of LLMs. Specifically, it involves three steps: modifying LLMs during the training phase, altering the logits generation during the inference phase, or adopting different token sampling strategies.

\textbf{Training Phase Embedding:} 
Embedding watermarks into LLMs during the training phase typically involves the use of backdoor techniques and data poisoning methods, as mentioned in Section~\ref{sec_generation}. This approach is inspired by backdoor attack~\cite{guBadNetsIdentifyingVulnerabilities2019a}, incorporating poisoned samples into the LLM's training data.
Assume the training data provider has his data samples $D_{\text{train}} = \{ (s_i, y_i) \}_{i=1}^{N}$ ,where each sample has its feature $s \in \mathcal{S}$ and label $y \in \mathcal{Y}$ .The attacker selects a small proportion of data $\{ (s_i, y_i) \}_{i=1}^{M}$, $M < N$ and adds a preset backdoor trigger to these samples while modifying their labels to a target label $\hat{y}$:
\begin{equation}
D_{\text{backdoor}} = \{ (s_i', \hat{y}) \}_{i=1}^{M}, s_i' = f_G(s_i, \text{trigger}),
\end{equation}
where $f_G$ is the watermark generation function to add \textit{trigger} into the input. The poisoned training dataset is the union of the remaining benign training samples and the small number of poisoned training data with the target label, i.e.,
\begin{equation}
D_{\text{Poisoned}} = D_{\text{train}} \cup D_{\text{backdoor}}.
\end{equation}

LLMs trained or fine-tuned using the dataset $D_{\text{Poisoned}}$ perform normally on original tasks but generate consistent, specific outputs when inputs contain a special trigger set $D_{\text{backdoor}}$. For example, Liu et al.~\cite{liuWatermarkingClassificationDataset2023} leveraged text backdoor techniques to insert triggers of different levels into a subset of the original training texts, uniformly changing the labels to a target label. Similarly, Sun et al.~\cite{sunCoProtectorProtectOpenSource2022} employed a similar data poisoning method to embed secret and stable watermark backdoors into open-source code. Modifying the labels of the training corpus can lead to a decrease in LLM performance. 
To mitigate the impact of changing labels, Tang et al.~\cite{tangDidYouTrain2023} proposed Clean-Label backdoor watermarking. After selecting the target category, adversarial perturbations are employed to ensure that the model learns features related to the backdoor while retaining the original labels.
Sun's CodeMark~\cite{sunCodeMarkImperceptibleWatermarking2023} introduces semantic-preserving transformations of code and builds poisoned training data by altering the syntactic form of the code, such as changing 'a+=1' to 'a=a+1'.

Methods that embed watermarks during the training of LLMs typically only provide a one-bit watermark information bit, which can only indicate the presence or absence of a watermark. Changes in the training process can lead to a decline in LLM performance and cause forgetting issues, thereby limiting the watermark embedding rate to a very low value. Additionally, altering the watermark requires retraining the LLM, restricting the application of watermarking algorithms that employ this embedding method.

During the training process, the embedded watermarks, whether introduced by adding subtle backdoor triggers or by embedding backdoors through semantic transformations, must ensure that the generated text $T^N$ remains within specified distortion constraints relative to the original text $S^N$ while maximizing the mutual information $I(S^N; T^N)$. Furthermore, without a trigger set, extracting the watermark $W$ from the watermarked text $T^N$ becomes significantly challenging, thereby ensuring the watermark's transparency.

\textbf{Inference Phase Embedding:} Watermark embedding in LLMs during the inference phase primarily diverges in two directions: modifying logits generation and employing different token sampling strategies. 

\textit{Logits Generation:}  Logits are scores assigned by the LLM to potential next words based on its internal representation and the input sequence, determining the probability distribution that influences the model's next word generation. Methods of watermark embedding at this stage includes all methods that guide logit modifications through vocabulary partitioning~\cite{kirchenbauerWatermarkLargeLanguage2023,kirchenbauerReliabilityWatermarksLarge2023,takezawaNecessarySufficientWatermark2023,huUnbiasedWatermarkLarge2023,quProvablyRobustMultibit2024,wuDiPmarkStealthyEfficient2023,fuWatermarkingConditionalText2024,leeWhoWroteThis2024,renRobustSemanticsbasedWatermark2024,wangCodableWatermarkingInjecting2023b,fernandezThreeBricksConsolidate2023,yooAdvancingIdentificationMultibit2024,zhaoProvableRobustWatermarking}, as well as various methods that produce watermark biases in logits~\cite{zhaoProtectingLanguageGeneration2023,liuUnforgeablePubliclyVerifiable2024,liuSemanticInvariantRobust2024,fairozePubliclyDetectableWatermarking2023},  directly inserting the watermark $W$ into the logits generated by LLMs. Essentially, these methods bias the logits or apply other methods to influence them, causing the LLM's output to exhibit a certain bias. Watermark embedding is achieved through this biased output. 

Assuming an LLM is trained on a vocabulary of size $V$, given a sequence of tokens as input, the LLM predicts the next token in the sequence by outputting a logit score vector $L_{ogit}$. Watermarks, represented by the red-green list~\cite{kirchenbauerWatermarkLargeLanguage2023} and generated through vocabulary partitioning, are embedded by modifying the logit scores:
\begin{equation}
    L_{ogit} = \begin{cases} 
L_{ogit} + \delta & \text{if watermarked}  \\
L_{ogit} & \text{otherwise}.
\end{cases}
\end{equation}

\begin{figure}[t]
    \centering
    \includegraphics[width=\linewidth]{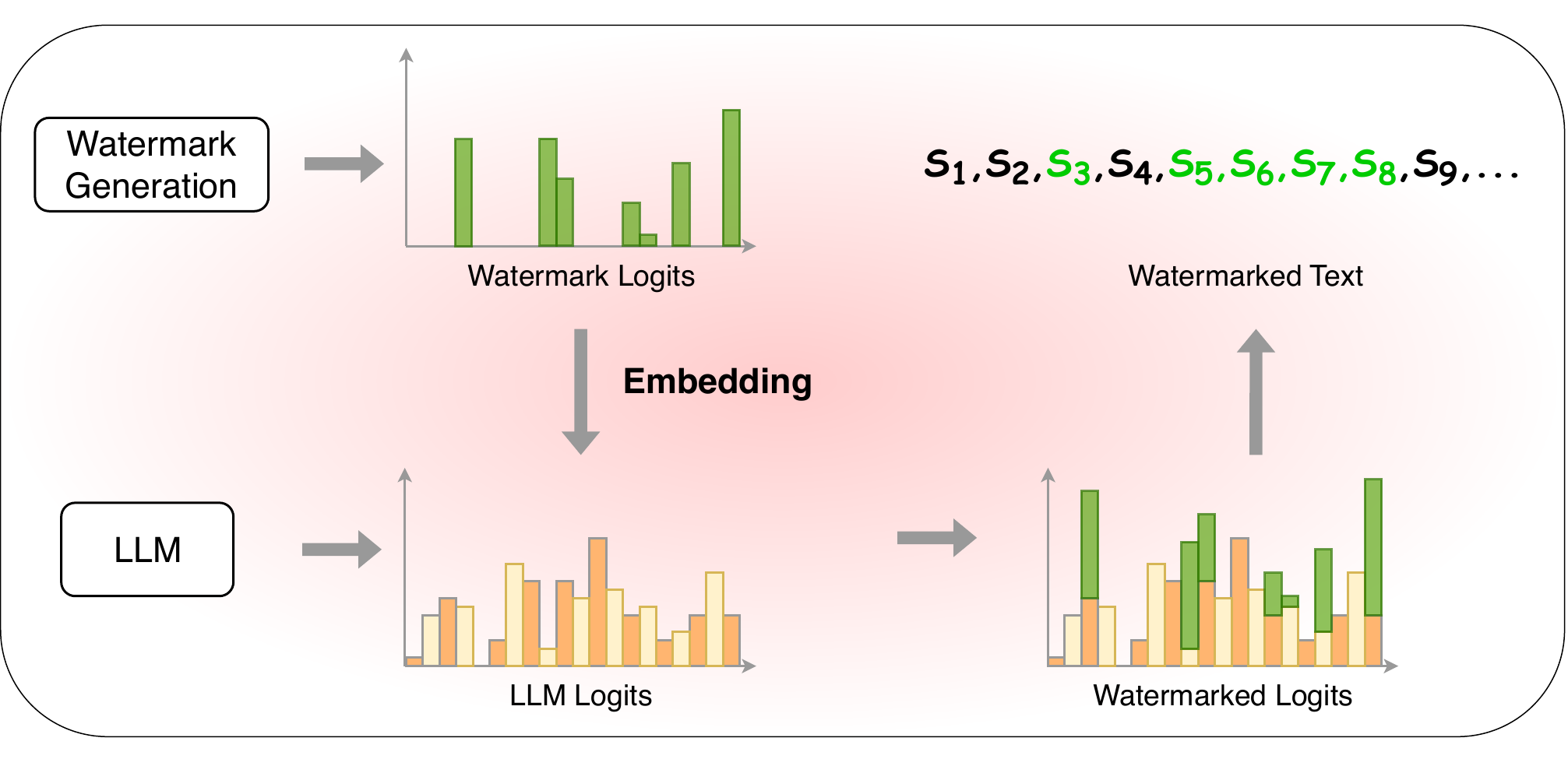}
    \caption{Watermark embedding through modifying logits generation.}
    \label{fig:logits_gen}
\end{figure}

As shown in Fig.~\ref{fig:logits_gen}, since the logits generated by the LLM have been modified, the LLM tends to select tokens from the generated watermark list, resulting in a higher proportion of the generated text being watermarked. In this way, the LLM is induced to exhibit a specific bias when selecting tokens, achieving the effect of embedding a watermark.

These embedding methods involve modifying the logits of LLMs to bias the model towards outputting tokens from a watermark list, considering the optimization target of Formula~\ref{eq6} during design. Kirchenbauer et al.~\cite{kirchenbauerWatermarkLargeLanguage2023} ensured the quality of the generated text $T^N$ by imposing constraints on the modified logits. Lee et al.~\cite{leeWhoWroteThis2024} ensured the functionality and quality of the code by avoiding embedding in low-entropy vocabulary. Takezawa et al.~\cite{takezawaNecessarySufficientWatermark2023} produced more natural texts than existing watermarking methods by adjusting the minimum constraints on logit modification based on the length of $S^N$. Furthermore, Hu et al.~\cite{huUnbiasedWatermarkLarge2023} defined two reweighting methods to produce an unbiased distribution of watermark logits, minimizing $I(W; T^N)$ to enhance the transparency of the watermark. DiPmark~\cite{wuDiPmarkStealthyEfficient2023} reweighted the watermark logits with a key, aiming to make $I(w; T^N) \to 0$ while maintaining the original distribution.

\textit{Token Sampling:} This section introduces embedding methods that intervene in the process of LLMs choosing the next token, utilizing the watermark W to guide the sampling strategy for each token to embed the watermark. Although the selection of tokens involves randomness, this randomness is controlled, with sampling strategies like random sampling, top-k, and top-p all possessing fixed randomness. By altering the sampling strategy with the watermark $W$, the watermark is embedded. During extraction, it is only necessary to judge the alignment between the chosen tokens and the set sampling sequence.

Christ et al.~\cite{christUndetectableWatermarksLanguage2023} use the output of a Pseudo-Random Function (PRF) to decide whether to embed a watermark at a specific location. Specifically, for each subsequent token generation decision, the LLM uses a bit (or a small part) of the PRF's output to determine whether to change that token to embed the watermark. If the output of the PRF is below a certain threshold in the model's original probability distribution, the model generates the token according to the original probability distribution; if it is above this threshold, the model will choose a different token to represent the watermark. Due to the fixed nature of pseudo-random numbers, the watermarked LLM will generate the same text for the same prompts every time, which limits the diversity of the text $T^N$. This corresponds to the second optimization target in Formula~\ref{eq6}, where $I(W; T^N)$ is larger, making it easier to find the relationship between $T^N$ and the watermark $W$ through multiple attack attempts.

To address the issue of monotonous generated text, Kuditipudi et al.~\cite{kuditipudiRobustDistortionfreeWatermarks2023} introduced the use of a random watermark key to compute a sequence of random numbers longer than the generated text, mapping it onto the sample to produce watermarked text. The alignment between the text and the pseudo-random number sequence is measured using the Levenshtein distance~\cite{yujian2007normalized}, thereby increasing the diversity of the text. Intervening in the sampling process of each token individually can easily lead to a decrease in text quality. Hou et al.~\cite{houSemStampSemanticWatermark2023} first calculated the LSH signature of the previously generated sentence, then randomly divided the LSH partitions into "valid regions" and "blocking regions" based on this signature. The watermark is embedded through the process of rejection sampling of tokens, meaning that new sentences are sampled from the language model until the embedding of the new sentence is located within the "valid region" of the semantic space, indicating successful embedding. Since the partitioning is based on the sentence level, this method significantly improves the quality of the generated text. Its performance in terms of watermark transparency and resistance to attacks far exceeds other methods.

\subsection{Watermark Extraction }

Watermark extraction methods in LLMs can be categorized into rule-based, trigger set-based, statistical, and deep learning-based approaches.

\subsubsection{Rule-based Methods}
Rule-based watermark extraction methods rely on identifying pre-defined patterns or rules within the text. These rules can include specific characters, combinations of words, text formatting, decision methods, password matching,  and other predefined conditions. By analyzing these features, rule-based methods detect suspected watermarked texts. These approaches pay special attention to specific indicators that suggest whether a text has been generated by AI, such as the entropy of the text, patterns of vocabulary usage, and sentence structure. In this context, entropy measures the randomness and complexity of the text, aiding in distinguishing between human and machine-generated texts. The primary advantage of rule-based methods lies in their simplicity and efficiency, as they do not depend on complex machine-learning models but instead rely on the direct analysis of specific text attributes. 

Some studies leverage the Quadratic Residues Theorem to formulate extraction rules for specific watermarking algorithms. Atallah et al.~\cite{atallahNaturalLanguageWatermarking2001} used a large prime as a key to calculate the hash values of the nodes in the semantic tree of each sentence,  converting them into binary strings. They then sorted the sentences based on these strings' hash values and extracted the watermark by reading specific bit sequences from each sentence following the marked sentences in the sorted sequence. Chiang et al.~\cite{chiangNaturalLanguageWatermarking2004} employed quadratic residue keys and bit operations to select terms from the text, constructed bit strings based on the values in the quadratic residue table, and ultimately transformed these bit strings using specific rules to extract the watermark. Topkara et al.~\cite{topkaraHidingVirtuesAmbiguity2006} relied on the weighted graph of synonym sets, using a secret key to select and color specific subgraphs. During watermark generation, words are replaced to embed watermark information, and during extraction, information is read based on a custom coloring scheme. Kim et al.~\cite{young-wonkimTextWatermarkingAlgorithm2003} extracted hidden information by calculating the statistical distribution of the space between words in text segments with the same category labels, using predefined decoding rules to extract from these statistical distributions. 

In the aforementioned studies, Zhao et al.~\cite{zhaoProtectingLanguageGeneration2023} employed the Lomb-Scargle periodogram method~\cite{scargle1982studies} to estimate the Fourier power spectrum. By applying an approximate Fourier transform, they amplified the subtle perturbations in the probability vector, enabling the detection of peaks in the power spectrum through frequency analysis to determine watermark information. He et al.~\cite{heCATERIntellectualProperty} developed the CATER method, which utilizes a set of features and relies on predefined conditions and discrimination rules to extract watermarks. Fairoze et al.~\cite{fairozePubliclyDetectableWatermarking2023}, on the other hand, calculated the hash value for each sequence using the watermark's public key and related parameters to determine whether it matches the expected signature to extract watermark information. These rule-based approaches are characterized by their specificity, making them unsuitable for generic watermark extraction. However, their advantage lies in the simplicity and intuitiveness of the extraction process.


\subsubsection{Trigger Set-based Methods}
The trigger set-based watermark extraction methods~\cite{liuWatermarkingClassificationDataset2023,tangDidYouTrain2023,sunCodeMarkImperceptibleWatermarking2023}
are typically used in conjunction with watermark embedding methods that employ backdoor techniques. The trigger sets usually consist of a group of backdoor text. 
Given a special trigger set, the LLMs will output a specific answer, which can be used as an extracted watermark. Since the implantation of backdoors participates in the LLM training process, these one-bit watermarks typically exhibit strong robustness. However, a key challenge of this approach lies in designing a covert backdoor trigger mechanism to ensure the transparency of the watermark and considering how to carry more information. Therefore, generating watermarks independent of the LLM training phase remains an area for further exploration.

\subsubsection{Statistical Methods}
Statistical watermark extraction methods involve rigorous mathematical and statistical analysis of texts or data to extract watermarks, focusing on identifying anomalies or characteristic differences in data distribution caused by watermark embedding. These approaches are particularly suited for detecting watermarks that have embedded statistical patterns or features during content generation. For instance, hidden watermarks can be extracted by comparing the statistical differences in vocabulary usage frequency, sentence length distribution, and syntactic complexity between watermarked and original texts. 

The extraction process analyzes the distribution of generated text tokens to determine if they follow the distribution introduced by the watermark. This is achieved using the Z-test, Likelihood Ratio Test, Q-offset detection, Jensen-Shannon Divergence, or other non-asymptotic statistical tests to determine whether the sample mean significantly deviates from its expected value. 

Works that generate watermarks based on vocabulary partitioning~\cite{kirchenbauerReliabilityWatermarksLarge2023,takezawaNecessarySufficientWatermark2023,huUnbiasedWatermarkLarge2023,quProvablyRobustMultibit2024,wuDiPmarkStealthyEfficient2023,fuWatermarkingConditionalText2024,leeWhoWroteThis2024,renRobustSemanticsbasedWatermark2024,wangCodableWatermarkingInjecting2023b,fernandezThreeBricksConsolidate2023,yooAdvancingIdentificationMultibit2024,zhaoProvableRobustWatermarking,pangAttackingLLMWatermarks2024a} and those based on model learning~\cite{houSemStampSemanticWatermark2023,kuditipudiRobustDistortionfreeWatermarks2023,liuSemanticInvariantRobust2024} used the Z-test for watermark extraction, similar to Kirchenbauer et al.~\cite{kirchenbauerWatermarkLargeLanguage2023}. Given a vocabulary partitioned into watermarked and non-watermarked tokens based on a fixed ratio, Z scores are calculated by computing the proportion of watermarked tokens to the total tokens. The null hypothesis is rejected, and the watermark is extracted from the text if the z-score exceeds a specified threshold. 

Hu et al.~\cite{huUnbiasedWatermarkLarge2023} proposed a watermark extraction method based on the log-likelihood ratio (LLR) score. This method calculates the LLR score for each watermarked text segment by comparing the relative probabilities of the text under two hypotheses. These scores are then aggregated, and the watermark is extracted if the aggregated LLR score exceeds a certain threshold. This method determines whether the text is more likely to originate from a distribution with a watermark.

SemaMark~\cite{renRobustSemanticsbasedWatermark2024} introduced Q-offset detection to enhance the robustness of boundary semantic values. This is achieved by searching for the highest z-statistic under different offset values Q, using it as the Q-offset score to correct variations in semantic values near boundaries. Li et al.~\cite{liProtectingIntellectualProperty2023} calculated the synonym distribution for each watermark channel and used the Jensen-Shannon Divergence (JSD) threshold to measure the similarity of these distributions to the original watermark distribution.

Statistical-based watermark extraction methods excel at handling large-scale text datasets and can be designed to be highly sensitive to minor changes in the data, achieving low false positive rates and false negative rates, thereby improving the accuracy of extraction. Additionally, statistical methods exhibit high robustness against watermark attacks, meaning that unless a large number of complex attack operations are carried out, it is difficult for attackers to remove the watermark without causing significant statistical deviations.

Despite the numerous advantages of statistical watermark extraction methods, they also have several notable disadvantages. One major drawback is their computational intensity, as the rigorous mathematical and statistical analysis required can be resource-demanding and time-consuming, particularly when handling large-scale text datasets. This makes them less suitable for real-time or near-real-time applications where speed is critical. Another issue is the dependency on the quality and characteristics of the input data; statistical methods may struggle with texts that lack sufficient statistical anomalies or differences for watermark detection, leading to potential inaccuracies. Furthermore, while these methods are generally robust against simple attacks, sophisticated adversaries with a deep understanding of the watermarking scheme can potentially devise complex strategies to manipulate the statistical properties of the data and evade detection. This continuous arms race between watermarking techniques and attack methods necessitates ongoing refinement and adaptation, posing a constant challenge for developers and researchers in the field.

\subsubsection{Deep Learning-based Methods}
Deep learning-based watermark extraction methods are often used in conjunction with watermark generation methods learned through models.  These approaches involve constructing and training deep neural networks to identify and extract hidden watermark patterns in text. They leverage pre-trained deep learning models, such as language models represented by encoders and decoders, to extract watermarks that are difficult to define directly through rules or statistical methods, utilizing the model's powerful feature extraction and pattern recognition capabilities. 

For instance, AWT~\cite{abdelnabiAdversarialWatermarkingTransformer2021} utilizes an adversarially trained watermark Transformer to extract watermark messages by automatically learning word replacements and positional information through a decoder. REMARK-LLM~\cite{zhangREMARKLLMRobustEfficient2023} extends AWT by using Transformers to predict inserted messages for watermark signature extraction. Yoo et al.~\cite{yooRobustMultibitNatural2023} utilize a pre-trained and fine-tuned filled model to identify masked positions based on text-invariant features, thereby extracting multi-bit watermark information. Liu et al.~\cite{liuUnforgeablePubliclyVerifiable2024} generate embeddings for all texts to be tested through an embedding network shared with the watermark generation network and then extract the watermark through binary classification using an LSTM network. Munyer et al.~\cite{munyerDeepTextMarkDeepLearningDriven2024} use the Bidirectional Encoder Representations from Transformers (BERT) pre-trained model as a powerful feature extractor for binary classification in watermark extraction, leveraging BERT's capability to capture the contextual meaning of words in a sentence. 

Wang et al.~\cite{wangCodableWatermarkingInjecting2023b} demonstrated that LLMs can serve as effective tools for watermark extraction,  showcasing significant potential in understanding and processing textual content. By leveraging the powerful language understanding capabilities of LLMs, they analyze subtle differences in vocabulary choice preferences, sentence structure variations, and syntactic complexity to extract specific watermark information. Existing watermark algorithms have largely overlooked the potential of LLMs as watermark extraction tools; future research could explore deeper integration between LLMs and robust semantic watermarks.

\subsection{Watermark Reconstruction}
The watermark reconstruction phase primarily targets multi-bit watermarks~\cite{zhangREMARKLLMRobustEfficient2023,liProtectingIntellectualProperty2023,wangCodableWatermarkingInjecting2023b,yooRobustMultibitNatural2023,fernandezThreeBricksConsolidate2023,fairozePubliclyDetectableWatermarking2023,yooAdvancingIdentificationMultibit2024,quProvablyRobustMultibit2024}, as one-bit watermarking algorithms can only verify the presence of a watermark during the extraction process. In contrast, multi-bit watermarks, which embed diverse customized messages, require the reconstruction of the customized message $m'$ from the watermark information space $\mathcal{M}$ based on the extracted watermark signal $W'$ after extraction. The reconstruction of watermark messages,  especially those identifying the source LLM, can be crucial for tracing misuse, such as spreading false information or academic dishonesty, back to the origin.

Wang et al.~\cite{wangCodableWatermarkingInjecting2023b}, based on the Maximum A Posteriori (MAP) decoding principle, designed a specific probability function $p$ in the reconstruction phase to measure the likelihood that the watermarked message is $m'$ given $W'$.

Some watermarking methods use reconstruction rules that correspond to those used during the watermark generation phase. For example, Li et al.~\cite{liProtectingIntellectualProperty2023} assigned a unique UID to each LLM user, with each bit of the UID representing a specific watermark channel. During reconstruction, they extracted the synonymous watermark tags from each watermark channel based on the UID and then sequentially reconstructed the watermark message for each bit according to the synonym substitution rules. Yoo et al.~\cite{yooAdvancingIdentificationMultibit2024} determined the position and color list of each token, incremented the token counts in the colored lists according to the division rules, and then determined the message content by identifying the color list with the most tokens at each message position.

Although some existing methods have successfully embedded multi-bit messages by providing different signals for each bit, multi-bit messages pose significant challenges during the reconstruction phase. As the bit width required for reconstruction grows exponentially with the increase in watermark information, maintaining the integrity of multi-bit watermark messages for reconstruction becomes increasingly difficult. Future research must further investigate the integrity of message reconstruction under noise and attack conditions. Additionally, the reconstruction phase must consider factors such as latency and computational cost, which significantly affect the user experience.

\subsection{Watermark Attacks}
With the development of LLM watermarking technology, corresponding methods for attacking these watermarks have also evolved, posing a significant threat to content security and copyright protection. These attack methods can be divided into four classes: destruction, extraction, forgery, and manipulation attacks. In this section, we comprehensively review the watermark attacking methods and assess their impact on LLM watermarking technology.

\begin{figure}[t]
    \centering
    \includegraphics[width=\linewidth]{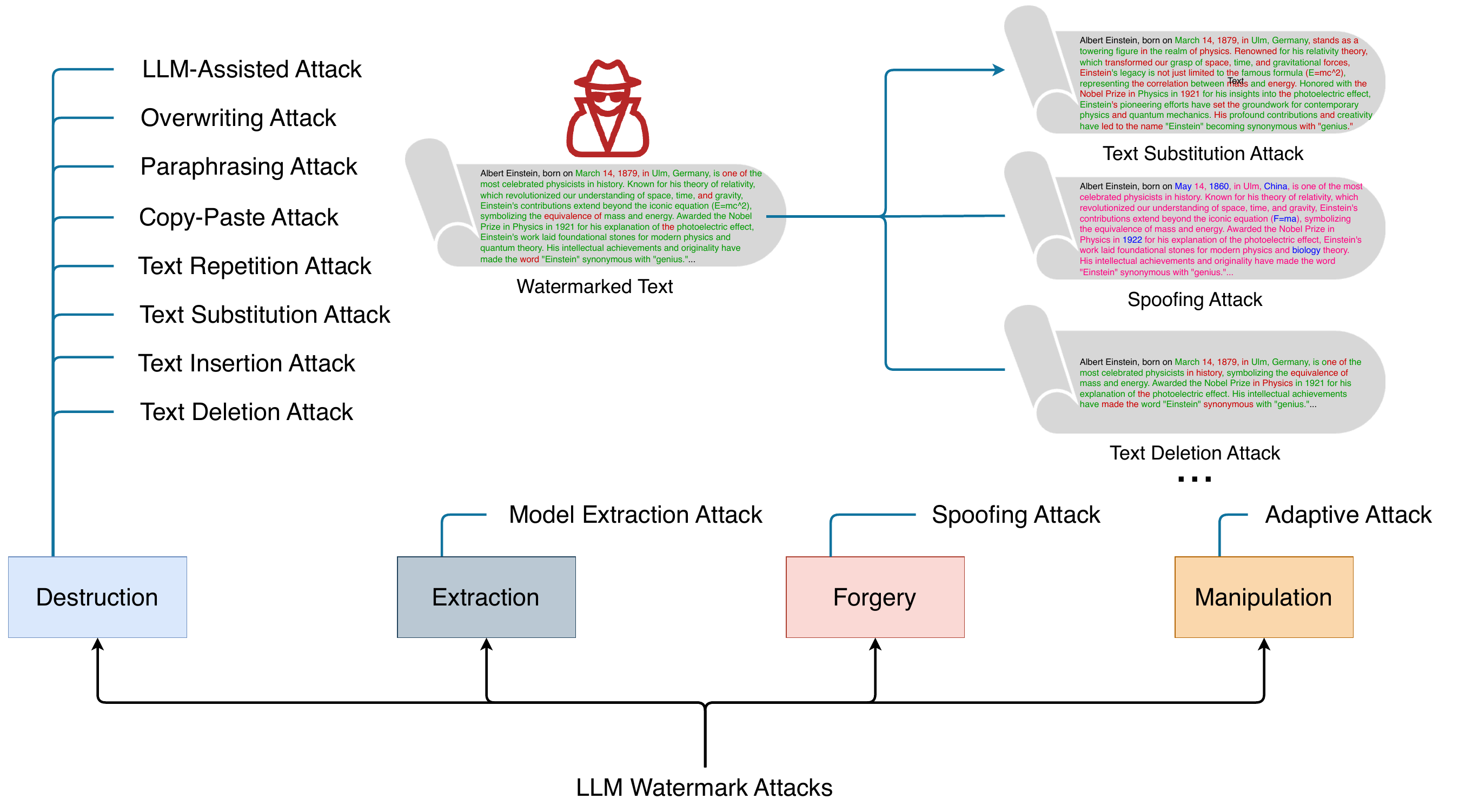}
    \caption{The overview of watermark attacks.}
    \label{fig:attack}
\end{figure}

Notably, these attacks not only represent potential threats but also serve as tools to assess the robustness of watermarking technologies, aiding developers in enhancing and fortifying watermark algorithms. Considering the intended objective of the attack, watermark attacks are categorized into four principal types: destruction, extraction, forgery, and manipulation, as shown in Fig.~\ref{fig:attack}.

\subsubsection{Destruction}
The primary purpose of these attacks is to destroy the watermark in text generated by LLMs, rendering it unextractable. Notably, removing a watermark is a trivial task if the quality of the language is disregarded. Therefore, the watermark attacks considered in this article follow Formula~\ref{eq9}, where attackers must operate under reasonable language quality constraints within a certain level of distortion. The attacks designed to compromise the watermark typically include text deletion, text insertion, text replacement, text repetition, copy-paste, paraphrasing, overwriting, and LLM-assisted attacks.

Text deletion attacks typically involve removing tokens from the generated text, and altering the watermark features by deleting tokens to increase the difficulty of watermark extraction. For example, Kirchenbauer et al.~\cite{kirchenbauerWatermarkLargeLanguage2023} have demonstrated that deleting tokens to remove green list tokens and modify the downstream red list can effectively destroy the watermark. However, this approach often significantly reduces the quality of the text. Some works have explored methods to enhance the robustness against text deletion attacks, such as introducing text edit distance as a soft constraint~\cite{kuditipudiRobustDistortionfreeWatermarks2023,zhaoProvableRobustWatermarking} and implementing a perturbation certification radius for changes in logits scores~\cite{wuDiPmarkStealthyEfficient2023}. These methods also enhance robustness against other types of text editing attacks, such as insertion and substitution. Moreover, text deletion attacks increase the cost of generation, as attackers "waste" generated tokens and significantly reduce the breadth of the LLM's context, explicitly lowering the text quality. Consequently, this level of distortion is usually intolerable.

Text insertion attacks involve adding extra tokens to the generated text to disrupt the watermark. Kirchenbauer et al.~\cite{kirchenbauerWatermarkLargeLanguage2023} demonstrated that inserting tokens from a red list can alter the calculation of the red list for downstream tokens. However, this modification changes the distribution of vocabulary, which poses a risk of reducing the quality of the text. Homoglyph attacks~\cite{gabrilovichHomographAttack2002} exploit the fact that Unicode characters are not unique, with multiple Unicode IDs resolving to the same or very similar letters. Boucher et al.~\cite{boucherBadCharactersImperceptible2021} found that injecting barely noticeable encodings, such as invisible characters or homoglyphs, can significantly degrade watermark performance. Overall, these types of attacks are difficult to detect visually but can be easily removed with various text formatting tools.

Text substitution attacks involve replacing one token with another specific token.  In~\cite{kirchenbauerWatermarkLargeLanguage2023}, tokens from a red list are introduced, increasing the proportion of downstream red list tokens. The homograph attack~\cite{gabrilovichHomographAttack2002} modifies text by substituting characters with identical or very similar ones. Building on this, Helfrich et al.~\cite{helfrichDualCanonicalizationAnswer2012} further formalized the attack. Works such as~\cite{leeWhoWroteThis2024} and~\cite{liProtectingIntellectualProperty2023} implement substitution by renaming code variables, while other studies~\cite{wangCodableWatermarkingInjecting2023b,yooRobustMultibitNatural2023,kuditipudiRobustDistortionfreeWatermarks2023,zhaoProvableRobustWatermarking,liuSemanticInvariantRobust2024} use synonyms for word replacement to assess the robustness of watermarks.

Text repetition attacks alter the original text's watermark distribution by repeating sections of text multiple times. This repetition affects the statistical tests used for watermark detection. Fernandez et al.~\cite{fernandezThreeBricksConsolidate2023} argue that human-generated texts with high repetition might be mistakenly labeled as machine-generated. The random variables used in watermarking methods, such as vocabulary partitioning, are only pseudo-random. Consequently, repetition produces the same patterns, altering the watermark distribution. This repetition undermines the assumption of independence required for calculating p-values.

The copy-paste attack~\cite{kirchenbauerReliabilityWatermarksLarge2023} has been employed by several  studies~\cite{liuSemanticInvariantRobust2024,yooAdvancingIdentificationMultibit2024,quProvablyRobustMultibit2024} to evaluate the robustness of watermarking. The principle involves mixing watermarked text segments generated by LLMs with manually written text, interspersing the watermarked portions within the surrounding unwatermarked text. Two controllable parameters in this attack are 1) the number of watermarked text segments inserted and 2) the proportion of the document containing watermarked text after the attack. This watermarking attack simulates a real-world scenario where attackers might not completely rewrite text generated by LLMs in practical applications but instead copy and paste it into a larger document to obscure its origin.  By employing it, researchers can assess whether watermarks remain detectable when the text is altered or combined with other texts.

Paraphrasing attacks, which involve rewriting text generated by LLMs using language models, human effort, or translation to maintain roughly the same meaning while employing different vocabulary choices and syntactic structures, significantly impact the robustness of watermarking. Specifically, the effectiveness of watermarking against such attacks depends on three factors: token sequence dependency, watermark strength, and text length. Krishna et al.~\cite{krishnaParaphrasingEvadesDetectors} demonstrated that rewriting LLM-generated text with a smaller language model can effectively evade existing AI-generated text detectors. They proposed two methods to enhance the effectiveness of paraphrasing attacks: context-aware rewriting of longer texts and increasing output diversity. Building on this, Sadasivan et al.~\cite{sadasivanCanAIGeneratedText2024} introduced the recursive paraphrasing attack, applying the paraphrasing process multiple times. After each iteration of paraphrasing, the resulting new text is re-entered into the paraphrasing model to generate further paraphrased text. Several studies~\cite{kirchenbauerReliabilityWatermarksLarge2023,yooAdvancingIdentificationMultibit2024,renRobustSemanticsbasedWatermark2024,liuSemanticInvariantRobust2024,liuUnforgeablePubliclyVerifiable2024} have designed language models with good paraphrasing performance for rewriting watermarked texts to assess the watermark's robustness against paraphrasing attacks. To improve resistance to paraphrasing attacks, Zhao et al.~\cite{zhaoProvableRobustWatermarking} employed a fixed vocabulary partitioning design to make the watermark less susceptible during paraphrasing. Hou et al.~\cite{houSemStampSemanticWatermark2023} utilized the semantic information of sentences to bolster robustness against paraphrasing attacks.

Watermark overwriting attacks involve regenerating the originally watermarked content or overwriting it with different watermarking methods. For instance, in REMARK-LLM~\cite{zhangREMARKLLMRobustEfficient2023}, new watermarks are used to rewrite the text in front of the original watermark, effectively circumventing the original watermark through the rewriting process.

Another category of attacks, referred to as LLM-assisted attacks, leverages the advanced capabilities of LLMs in understanding and generating human language to conduct attacks. For example, the Goodside emoji attack~\cite{goodside2023adversarial} as discussed in~\cite{kirchenbauerWatermarkLargeLanguage2023,christUndetectableWatermarksLanguage2023} involves instructing the model to produce responses that prompt the insertion of emojis between every pair of words. This type of attack disrupts any watermark that relies on the watermark extractor seeing a continuous sequence of tokens. Consequently, vocabulary partitioning methods generated from the previous discussion cannot resist such attacks.

\subsubsection{Extraction}
The goal of model extraction attacks is to imitate the behavior of a target model, creating a valuable local model to evade substantial service fees or even to offer competitive services. In such attacks, attackers create an unwatermarked copy by replicating or mimicking the functionality of the protected model. This type of attack specifically targets scenarios where watermarks are embedded into model outputs to claim intellectual property rights. Attackers may make numerous queries to understand and replicate the model's behavior, thereby constructing a similar-performing model copy that does not contain the watermark. Li et al.~\cite{liProtectingIntellectualProperty2023} discussed the feasibility of implementing extraction attacks through LLM APIs. The presence of watermarks in the stolen high-quality LLM imitation models suggests that the proposed watermark is both invisible and robust.

\subsubsection{Forgery}
Watermark spoofing attacks refer to attackers modifying or constructing text so that clean text without any watermark is incorrectly identified by the watermark extractor as containing a legitimate watermark, or causing the extractor to return incorrect watermark information from the victim organization. These attacks exploit flaws in the watermark algorithm's extraction and detection mechanisms, particularly when the extraction process relies on rules or statistics.

Here are some examples of watermark spoofing attacks: 1. Attackers can leverage spoofing attacks to fabricate fake news or misinformation and publish it on public media, falsely claiming through manipulated watermarks that the fake news was produced by a legitimate company's LLM. 2. Attackers can embed a benign company's watermark within malicious code using spoofing attacks, making the benign company responsible for the harm caused by the malicious code.

Some works have explored spoofing attacks on watermarked LLMs. For instance, Sadasivan et al.~\cite{sadasivanCanAIGeneratedText2024} artificially constructed text with an understanding of the watermarking method, leading the extractor to misjudge the presence of a watermark. Nevertheless, their approach requires an excessive number of queries from the attacker (1 million), limiting its applicability to only the KGW~\cite{kirchenbauerWatermarkLargeLanguage2023} watermarking scheme, thus making it difficult to generalize to other watermarks. ~\cite{guLearnabilityWatermarksLanguage2024} trained a novel model to learn the distribution of watermarked tokens, which is not feasible for attackers with limited computational resources, especially due to the substantial requirement for a multitude of queries to construct training data and to train a new LLM. Pang et al.~\cite{pangAttackingLLMWatermarks2024a} argued that robust watermarks may need to compromise on robustness to mitigate the possibility of spoofing attacks. The robustness of LLM watermarks reduces the difficulty of spoofing attacks, as attackers do not need to ensure that every modification or misleading token is watermarked; they only need the overall detection confidence score to exceed a threshold to consider the text content as generated by a watermarked LLM. To address this, Liu et al.~\cite{liuSemanticInvariantRobust2024} incorporated watermarking rules intertwined with textual semantic information, proving to be an effective method to withstand spoofing attacks.

\subsubsection{Manipulation}

Attackers often utilize adaptive attack techniques to achieve control and manipulation of watermarking algorithms. These attacks are highly customized, assuming that the attacker possesses knowledge of either the entire or partial watermarking framework. With this insight, they can tailor specific adjustments and optimizations based on the characteristics and detection mechanisms of the watermark, thereby increasing the success rate of their attack. Moreover, armed with insider information about the watermark, they can effortlessly produce or replicate private watermarks. Li et al.~\cite{liProtectingIntellectualProperty2023} evaluated the reliability of watermarks under strong adaptive attacks, investigating whether attackers could manipulate text to make watermark extraction fail when they understand the principles of watermark embedding. Liu et al.~\cite{liuUnforgeablePubliclyVerifiable2024} found that even in scenarios where attackers have access to the watermark extractor and can make unlimited queries to understand the watermark generation rules, it remains difficult to infer the watermark generation method. Existing research rarely uses adaptive attacks for the evaluation of watermark performance, indicating a need for further exploration of attackers' capabilities to conduct a more intricate analysis of the resilience and unforgeability of watermarks.

\section{Evaluation Metrics}
\label{sec4}
\begin{figure}[H]
    \centering
    \includegraphics[width=\linewidth]{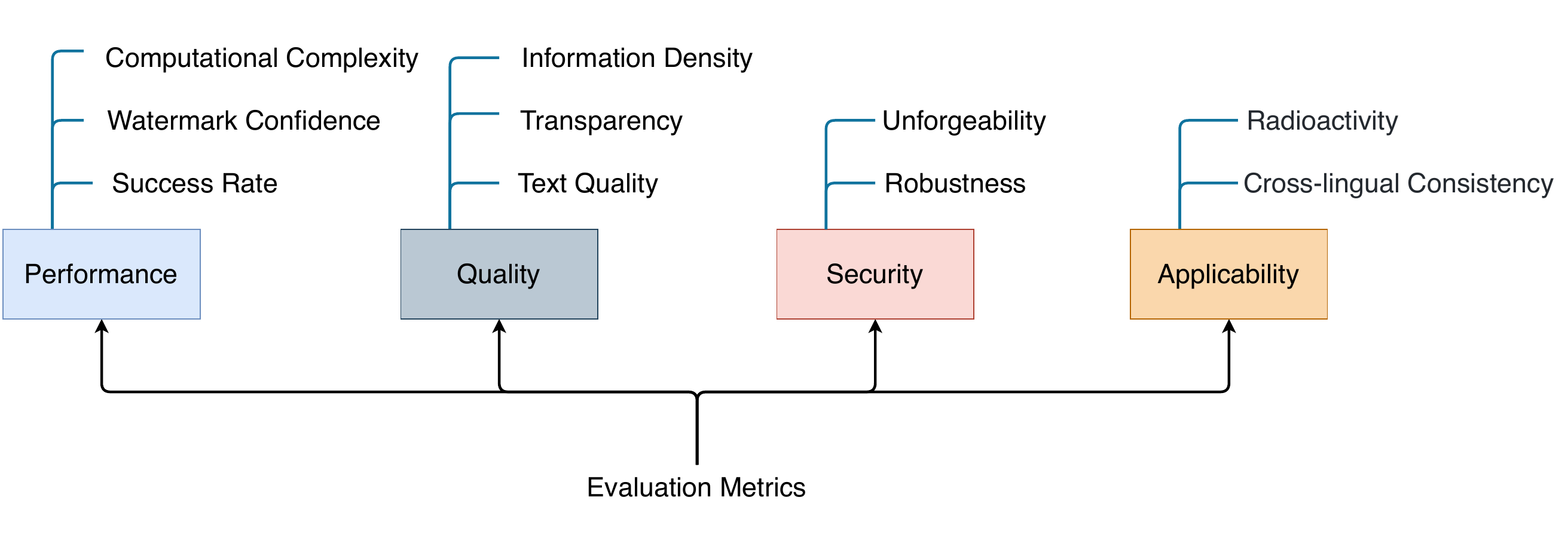}
    \caption{The categorization of watermark evaluation metrics.}
    \label{fig:metrics}
\end{figure}

A comprehensive and standardized evaluation system is essential for watermarking algorithms in LLMs. As shown in Fig.~\ref{fig:metrics}, this section outlines the evaluation metrics for LLM watermark algorithms from four perspectives: performance, quality, security, and applicability. These metrics include success rate, watermark confidence, computational complexity, text quality, transparency, information density, robustness, unforgeability, cross-language consistency, and radioactivity. A detailed summary of these metrics aids in thoroughly understanding the effectiveness of watermark algorithms, thereby guiding future research directions and practical applications of LLM watermarks. Furthermore, the evaluation system established upon these metrics assists researchers in selecting or designing watermark algorithms and plays a crucial role in assessing their feasibility and effectiveness in real-world scenarios. Finally, we summarize the varied emphases on watermark algorithm requirements among four entities in LLM applications and the related watermarking methods for different entities based on their focus metrics in Table~\ref{tab2}.

\begin{table}[t]
  \centering
  \caption{Relationships between LLM entities and watermarking algorithm requirements, and a list of related watermark algorithms. $\star$ stands for basic requirements, $\bullet$ stands for primary requirements, and $\circ$ stands for secondary requirements.}
    \begin{tabular}{ccccc}
    \toprule
    \multirow{2}[4]{*}{\textbf{Metrics}} & \multicolumn{4}{c}{\textbf{Entities}} \\
\cmidrule{2-5}          & \multicolumn{1}{c}{Data Provider} & \multicolumn{1}{c}{Technology Provider} & \multicolumn{1}{c}{LLM User} & \multicolumn{1}{c}{PRTTP} \\
    \midrule
    Success Rate &    $\star$   &   $\star$    &   $\star$    & $\star$ \\
    Text Quality &  $\star$     &   $\bullet$    &   $\circ$    &  $\circ$ \\
    Watermark Confidence &  $\bullet$     &   $\bullet$    &   $\bullet$    & $\star$ \\
    Robustness &   $\star$    &   $\star$    &     $\circ$  & $\bullet$ \\
    Unforgeability &  $\bullet$     &    $\star$   &  $\bullet$     & $\star$ \\
    Transparency &  $\bullet$     &  $\bullet$     &   $\circ$     & $\circ$  \\
    Information Density &  $\bullet$     &  $\bullet$     &  $\bullet$     & $\bullet$ \\
    Computational Complexity &  $\bullet$     &    $\star$   &   $\star$    &  $\circ$ \\
    Cross-lingual Consistency &   $\star$     &    $\star$   &   $\circ$    & $\bullet$ \\
    Radioactivity &   $\star$    &  $\bullet$     &    $\circ$   & $\circ$ \\
    \midrule
    \multirow{3}[2]{*}{Related Methods} & \multicolumn{1}{l}{\cite{kirchenbauerReliabilityWatermarksLarge2023},\cite{sunCodeMarkImperceptibleWatermarking2023},\cite{wangCodableWatermarkingInjecting2023b}}  & \multicolumn{1}{l}{\cite{wangCodableWatermarkingInjecting2023b},\cite{wuDiPmarkStealthyEfficient2023},\cite{takezawaNecessarySufficientWatermark2023},\cite{yooAdvancingIdentificationMultibit2024},\cite{leeWhoWroteThis2024}}& \multicolumn{1}{l}{\cite{wangCodableWatermarkingInjecting2023b},\cite{kuditipudiRobustDistortionfreeWatermarks2023},\cite{yangWatermarkingTextGenerated2023}} &\multicolumn{1}{l}{\cite{fairozePubliclyDetectableWatermarking2023},\cite{liuUnforgeablePubliclyVerifiable2024}} \\
    & \multicolumn{1}{l}{\cite{kuditipudiRobustDistortionfreeWatermarks2023},\cite{liuSemanticInvariantRobust2024},\cite{renRobustSemanticsbasedWatermark2024}}&\multicolumn{1}{l}{\cite{fernandezThreeBricksConsolidate2023},\cite{huUnbiasedWatermarkLarge2023},\cite{liuSemanticInvariantRobust2024},\cite{renRobustSemanticsbasedWatermark2024},\cite{quProvablyRobustMultibit2024}} &\multicolumn{1}{l}{\cite{liProtectingIntellectualProperty2023},\cite{christUndetectableWatermarksLanguage2023},\cite{munyerDeepTextMarkDeepLearningDriven2024}} & \multicolumn{1}{l}{\cite{kuditipudiRobustDistortionfreeWatermarks2023},\cite{liuSemanticInvariantRobust2024}} \\
    & \multicolumn{1}{l}{\cite{yooRobustMultibitNatural2023},\cite{tangDidYouTrain2023},\cite{munyerDeepTextMarkDeepLearningDriven2024} }& & &\multicolumn{1}{l}{\cite{wuDiPmarkStealthyEfficient2023}}  \\

    \bottomrule
    \end{tabular}%
  \label{tab2}%
\end{table}%

\subsection{Success Rate}
The success rate serves as the primary metric for assessing the effectiveness of watermarking technologies, directly reflecting the ability of the watermark extraction phase to identify or extract messages embedded with watermarks accurately~\cite{liuSurveyTextWatermarking2024}. 

\textbf{One-bit Watermark:} For one-bit watermarks, the success rate is typically assessed through metrics such as the precision of the extraction algorithm, AUROC (Area Under the Receiver Operating Characteristic curve), and F1 score.

Accuracy, which measures the proportion of correctly detected watermarked texts among all identified texts, is not commonly employed to evaluate watermark performance due to the highly imbalanced distribution of datasets in the context of text watermark detection—where only a minority of texts carry watermarks. In such cases, models could achieve high accuracy by indiscriminately labelling all texts as non-watermarked, thereby failing to truly capture the actual efficacy of the watermarks.

Some studies~\cite{houSemStampSemanticWatermark2023,zhaoProvableRobustWatermarking,renRobustSemanticsbasedWatermark2024,yooAdvancingIdentificationMultibit2024} use AUROC, which measures the model's overall effectiveness in distinguishing between watermarked and non-watermarked texts across different thresholds. This provides a more stable and reliable metric for specific performance evaluation tasks in watermarking, as it reflects the model's average performance across all possible classification thresholds without being directly influenced by any single threshold setting.

Additionally, some studies~\cite{zhaoProtectingLanguageGeneration2023,yooAdvancingIdentificationMultibit2024,liProtectingIntellectualProperty2023} employ precision as a metric. Precision directly reflects the model's efficacy in identifying watermarked texts, particularly in applications focused on the minimization of misclassifications. It measures the proportion of samples predicted as positive that are indeed positive. In text watermark detection, precision represents the proportion of texts correctly identified as containing watermarks out of all texts marked as containing watermarks. 

The F1 score, a harmonic mean of precision and recall (the proportion of actual watermarked texts correctly identified), balances the comprehensive detection performance. It is widely employed in the evaluation of watermark performance in studiess~\cite{kirchenbauerWatermarkLargeLanguage2023,liuUnforgeablePubliclyVerifiable2024,liuSemanticInvariantRobust2024,zhaoProvableRobustWatermarking,renRobustSemanticsbasedWatermark2024,quProvablyRobustMultibit2024}.

\textbf{Multi-bit Watermark:} For informative watermarks, the success rate reflects the ability to correctly identify and recover the watermarked information after potential attacks or disruptions. In such cases, the success rate is refined to evaluate the percentage of successfully extracted watermarked messages during the watermark extraction and message reconstruction stages. Zhang et al.~\cite{zhangREMARKLLMRobustEfficient2023} and Wang et al.~\cite{wangCodableWatermarkingInjecting2023b} calculate the ratio of successfully recovered messages to the total number of messages. Meanwhile, Yoo et al.~\cite{yooRobustMultibitNatural2023} use the Bit Error Rate (BER) to calculate the ratio of incorrectly recovered message bits to the total bit count, providing a more detailed assessment of the watermark extraction success rate.

Some studies~\cite{abdelnabiAdversarialWatermarkingTransformer2021,yooAdvancingIdentificationMultibit2024} utilize bit accuracy, which denotes the rate of correct bit predictions. Essentially, it is similar to the Bit Error Rate (BER). The information density of the watermarking algorithm significantly influences the success rate of existing multi-bit watermarks. For watermarks with a limited information payload, increasing the information density can easily reduce the success rate of watermark extraction~\cite{liuSurveyTextWatermarking2024}.

\subsection{Watermark Confidence}
Watermark confidence serves as a metric utilized to measure the reliability and certainty of watermark information embedded in text, reflecting the credibility of watermark detection results. Confidence is measured through statistical tests to assess the credibility of watermark detection outcomes, helping to determine whether the detected watermark has statistical significance. In text watermarking technology, high confidence means that we can be very sure that the detected watermark truly exists, rather than being a false positive result caused by random noise or error. The assessment of watermark confidence stands as a pivotal task in ensuring the practicality and effectiveness of watermarking technology, especially in fields like copyright protection, where accurately extracting and verifying watermark information is necessary.

Z-Score (standard score) is a commonly used measure in statistics, indicating the deviation of a value from the mean of its dataset in terms of standard deviation units. In the context of watermark confidence, Z-Score can be used to measure the deviation of the detected watermark signal strength from the standard deviation of background noise. The formula is:

\begin{equation} 
Z = \frac{X - \mu}{\sigma} 
\end{equation}

\noindent where $X$ is the observed watermark signal strength, $\mu$ is the mean strength of the background noise, and $\sigma$ is the standard deviation of the background noise strength. A high Z-Score value indicates that the watermark signal significantly surpasses or falls below the average background noise level, consequently increasing confidence in the presence of the watermark.

P-value measures the probability of observing the data under the condition that the null hypothesis is true. In the context of text watermarking, the null hypothesis typically states: "The detected signal is merely random noise, and no watermark is present."
A lower P-value indicates stronger support for rejecting the null hypothesis, thereby enhancing our confidence in the actual existence of the watermark. In statistical testing, if the P-value falls below a pre-set significance threshold (e.g., 0.05~\cite{fisher1966design}), then the result is considered statistically noteworthy,  implying high confidence regarding the watermark's existence.

Certain algorithms that extract watermarks using statistical methods employ Z-Score~\cite{kirchenbauerWatermarkLargeLanguage2023,kirchenbauerReliabilityWatermarksLarge2023,yangWatermarkingTextGenerated2023,zhangREMARKLLMRobustEfficient2023,takezawaNecessarySufficientWatermark2023,huUnbiasedWatermarkLarge2023,zhaoProvableRobustWatermarking,wuDiPmarkStealthyEfficient2023,yooAdvancingIdentificationMultibit2024,fuWatermarkingConditionalText2024,renRobustSemanticsbasedWatermark2024,liuSemanticInvariantRobust2024,liuUnforgeablePubliclyVerifiable2024,leeWhoWroteThis2024,pangAttackingLLMWatermarks2024a} or P-Value~\cite{abdelnabiAdversarialWatermarkingTransformer2021,heCATERIntellectualProperty,kuditipudiRobustDistortionfreeWatermarks2023,sunCodeMarkImperceptibleWatermarking2023,fernandezThreeBricksConsolidate2023} to assess the confidence level of the watermark. When the confidence level surpasses a predefined threshold, the hypothesis that a watermark exists is considered valid.
In practice, the Z-Score and P-Value are often used in conjunction to assess the confidence level associated with a watermark. This approach quantifies the strength of the watermark signal and provides statistical evidence to support its existence. First, the Z-Score is calculated to quantify the importance of the detected watermark signal in relation to the background noise. Subsequently, the corresponding P-value is calculated utilizing the Z-Score to determine whether this significance reaches a statistically significant level. By calculating the confidence level, researchers can more accurately assess the effectiveness and reliability of the watermark.

\subsection{Computational Complexity}
Computational complexity focuses on the time and resources consumed during the generation and extraction phases of watermarking. It can be evaluated by directly measuring the actual generation, embedding, extraction, and reconstruction times and the computational resources required to perform these operations. 

Some works~\cite{zhangREMARKLLMRobustEfficient2023,yooAdvancingIdentificationMultibit2024} discuss the processing time and memory overhead of generating watermarks, while Lee et al.~\cite{leeWhoWroteThis2024} test detection times using proxy detection models of various sizes. Wang et al.~\cite{wangCodableWatermarkingInjecting2023b} explore the time required for generation and extraction under different watermark settings, as well as the time consumed by different proxy LLMs in watermark generation. Hou et al.~\cite{houSemStampSemanticWatermark2023} utilize parallel rejection sampling to reduce the time taken to generate watermark texts. Fairoze et al.~\cite{fairozePubliclyDetectableWatermarking2023} theoretically derive the computational cost of asymmetric private key signatures.

The level of computational complexity directly affects the practicality and feasibility of watermarking techniques. When designing and evaluating text watermarking schemes, computational complexity is an indispensable factor. An ideal watermarking scheme should minimize the time and resource consumption of the encoding and decoding processes while ensuring the watermark's concealability, robustness, and capacity.

\subsection{Text Quality}
LLM watermarking techniques should embed watermarks without significantly impacting the original text's quality. Text quality is commonly measured using metrics such as perplexity and semantic scores. A low perplexity indicates high text coherence and readability, indicating more accurate model predictions. Semantic scores evaluate the semantic consistency between the watermarked text and the original, ensuring the meaning of the text remains unchanged after watermark embedding. In practice, natural language processing technologies assess semantic preservation by computing the cosine similarity of text embedding vectors. Additionally, a comprehensive evaluation of the generation text's Perplexity, Semantic Score, BLUE, Rouge, Edit Distance, and other metrics, combined with specific downstream tasks of the dataset, can be conducted.

\textbf{Perplexity (PPL)} is an indicator used to measure the smoothness of the probability distribution predicted by language models.  It is a valuable tool for evaluating the consistency and fluency of text. A low perplexity indicates the probability distribution is adept at predicting the given sample. In the context of text watermarking, optimizing the PPL of watermarked texts can help ensure that the embedding of watermarks does not disrupt the text's fluency and coherence. Specifically, given a text sequence $S^N = (s_1 \ldots s_N)$, the perplexity (PPL) can be computed using an LLM as
\begin{equation}
    PPL(S|\text{Prompt}) = \left[ \prod_{i=1}^{N} P_{LLM}(s_i|\text{Prompt}, s_{(i-1)}) \right]^{-\frac{1}{N}}.
\end{equation}

Calculating PPL using Oracle LLMs with a larger number of parameters and stronger semantic capabilities can yield a more accurate evaluation, such as with models like GPT-2~\cite{yangWatermarkingTextGenerated2023,yooAdvancingIdentificationMultibit2024}, GPT-3~\cite{zhaoProvableRobustWatermarking,pangAttackingLLMWatermarks2024a,quProvablyRobustMultibit2024}, OPT-2.7B~\cite{kirchenbauerWatermarkLargeLanguage2023,wangCodableWatermarkingInjecting2023b,houSemStampSemanticWatermark2023,renRobustSemanticsbasedWatermark2024}, LLaMA-7B~\cite{takezawaNecessarySufficientWatermark2023,wuDiPmarkStealthyEfficient2023}, LLaMA-13B~\cite{liuUnforgeablePubliclyVerifiable2024,liuSemanticInvariantRobust2024}, etc. Generally speaking, the goal is to maintain consistency in Perplexity (PPL) between the watermarked text and the original text when assessed on the same oracle LLM. This alignment helps guarantee that the process of watermarking does not result in a noticeable decline in the quality of the text.

\textbf{Semantic Scores} reflect the semantic similarity between the watermarked text and the original text. Evaluating semantic scores typically involves employing language models to calculate the semantic embeddings of sequences and then comparing these embeddings through cosine similarity. Semantic similarity is evaluated in various ways, including \textbf{BERTScore}~\cite{zhangBERTScoreEvaluatingText2020} and \textbf{GPTScore}~\cite{fuGPTScoreEvaluateYou2023}. Some studies \cite{zhaoProtectingLanguageGeneration2023,heCATERIntellectualProperty,zhangREMARKLLMRobustEfficient2023,huUnbiasedWatermarkLarge2023,houSemStampSemanticWatermark2023,wuDiPmarkStealthyEfficient2023,fernandezThreeBricksConsolidate2023} utilize BERTScore to calculate the similarity scores for each token in the candidate sentence with every token in the reference sentence, focusing on semantic equivalence through contextual embeddings. Hu et al.~\cite{huUnbiasedWatermarkLarge2023} use GPTScore to score by calculating the conditional probability of generating a specific text under a given context and evaluation protocol and utilize the full potential of pre-trained models for text evaluation. Semantic scores help measure the semantic similarity between watermarked texts and original texts. Approaching from a semantic perspective, a more precise and detailed exploration of the complex semantic relationships among texts can assist in developing semantic watermarks with stronger robustness against text-editing attacks.

\textbf{BLEU (Bilingual Evaluation Understudy)}~\cite{papineniBleuMethodAutomatic2002} is commonly used in the machine translation domain to assess translation quality by comparing the n-gram overlap between machine translation outputs and a set of reference translations. In text watermark scenarios, BLEU can measure the lexical similarity between watermarked texts and original texts, thus ensuring the naturalness of language and preservation of original intent. Some studies~\cite{zhaoProtectingLanguageGeneration2023,heCATERIntellectualProperty,zhangREMARKLLMRobustEfficient2023,sunCodeMarkImperceptibleWatermarking2023,liProtectingIntellectualProperty2023,takezawaNecessarySufficientWatermark2023,fuWatermarkingConditionalText2024}  compare the translation outputs of watermarked LLMs with those of the original LLMs and find that watermarking leads to a reduction in BLEU scores. To address this, the studies conducted by Hu et al.~\cite{huUnbiasedWatermarkLarge2023} and Wu et al.~\cite{wuDiPmarkStealthyEfficient2023} propose the use of unbiased watermarks to maintain BLEU scores, ensuring the quality of text translation.
Compared to the BLEU score, the \textbf{METEOR score}~\cite{denkowskiMeteorUniversalLanguage2014} is a more advanced metric for assessing translation quality. It also compares machine translation outputs with reference translations but takes additional information such as synonyms, word forms, and sentence structure into account. Yang et al.~\cite{yangWatermarkingTextGenerated2023} replace BLEU with METEOR score to evaluate the quality of watermarked texts.

Some studies~\cite{zhaoProtectingLanguageGeneration2023,huUnbiasedWatermarkLarge2023,wuDiPmarkStealthyEfficient2023,fuWatermarkingConditionalText2024} employ \textbf{ROUGE}~\cite{linROUGEPackageAutomatic2004}  to automatically determine the quality of text by comparing watermarked texts with other (ideal) human-created texts. To evaluate whether watermarking operations affect the core information and quality of summaries, ROUGE calculates the count of overlapping units, such as n-grams, word sequences, and word pairs, between the watermarked texts generated by the LLM being evaluated and the ideal texts created by humans. 

In addition to these common text quality assessment metrics, there exist other metrics used to explore the impact of watermarking algorithms on specific domain tasks. For instance, Lee et al. ~\cite{leeWhoWroteThis2024} utilize \textbf{Code Quality Pass@k} to measure the pass rate of code snippets generated by LLMs under given test cases. Maintaining a high pass rate for code embedded with watermarks is crucial to ensuring the functionality of the code remains unaffected. Zhao et al.~\cite{zhaoProvableRobustWatermarking} use \textbf{Edit Distance} to measure sequence differences, quantifying the extent of changes after text editing. Yoo et al.~\cite{yooAdvancingIdentificationMultibit2024} employ\textbf{ P-SP }(Semantic Similarity based on Paraphrase Model)~\cite{wietingParaphrasticRepresentationsScale2023} to measure the semantic similarity between human texts and watermarked texts given the same prompts. For the assessment of text diversity, Hou et al.~\cite{houSemStampSemanticWatermark2023} propose two metrics:\textbf{ Ent-3} and \textbf{Rep-3}. Ent-3 achieves this by calculating the entropy of the frequency distribution of trigrams in the generated text. A higher Ent-3 value indicates the text has greater diversity. Conversely, the Rep-3 metric measures the proportion of repeated trigrams in the generated text. A lower Rep-3 value indicates less repetition in the text, thereby enhancing the text's diversity and quality.

No single metric comprehensively covers all aspects of quality evaluation. Consequently, within the realm of text quality assessment, a multifaceted approach is imperative. This involves considering a broad spectrum of criteria, such as output dispersion, semantic integrity, task-specific textual fluency, and diversity of the text. A comprehensive evaluation of text quality, therefore, demands the integration of multiple metrics to gauge the overall performance and quality of the text accurately. By adopting this comprehensive approach, a more nuanced and comprehensive evaluation emerges, aligning with the multifaceted essence of text quality within watermarking research.

\subsection{Transparency}
Watermark transparency is a critical metric strongly related to text quality, used to assess the indistinguishability of watermarked text from the original text, both visually and statistically. This attribute reflects the watermark's ability to remain undetected. Even under meticulous scrutiny, the presence of the watermark is not readily apparent. This ensures the confidentiality of the watermark messages and maintains the naturalness of the original text. The evaluation of transparency typically relies on human assessment or machine learning models to test whether the watermark can be extracted from the text. This phenomenon is quantified through the false positive rate (incorrectly identifying an unwatermarked text as watermarked) and miss rate (failing to identify watermarked text).

Optimizing watermark transparency necessitates consideration across multiple dimensions, including visual indistinguishability~\cite{heProtectingIntellectualProperty2022,munyerDeepTextMarkDeepLearningDriven2024}, consistency of statistical properties~\cite{christUndetectableWatermarksLanguage2023,huUnbiasedWatermarkLarge2023}, and semantic alignment~\cite{houSemStampSemanticWatermark2023,liuSemanticInvariantRobust2024,renRobustSemanticsbasedWatermark2024}. By employing intricately designed watermarking schemes, it is feasible to effectively conceal watermark information without compromising the naturalness and readability of the text.

\subsection{Information Density}
Information density is a critical concept in the field of text watermarking. High-density watermarking algorithms can carry more identifiers and copyright information within the same amount of text, thereby enhancing identification. 

Information density can be characterized as the amount of information embedded per unit of text length (e.g., word, sentence, or paragraph). This measure, also known as payload, can be calculated using Shannon entropy. The calculation method involves analyzing the number of symbols in the text available for encoding and their probability distribution, thereby determining the maximum possible information density. This concept is crucial in designing and evaluating text watermarking schemes because it directly affects the watermark's concealability, robustness, and capacity.

We define empirical entropy, as a measure used to estimate the information density of a system, reflecting its uncertainty or randomness. The formula for Shannon entropy H(X) for a discrete random variable X with possible values $({x_1, x_2, ..., x_n})$ and probability mass function P(X) is given by

\begin{equation}
    H(X) = -\sum_{i=1}^{n} P(x_i) \log_b P(x_i),
\end{equation}

\noindent where $P(x_i)$ is the probability of occurrence of the value $x_i$, and $b$ is the base of the logarithm, commonly set to 2 for binary systems, resulting in units of bits.

In the context of text watermarking, the symbols used for encoding the watermark (e.g., variations in word choice, syntax, or punctuation) represent the "symbols" in the Shannon entropy formula. The probability distribution of these symbols depends on how frequently they can be used for watermarking without altering the natural flow or meaning of the text. Amplified entropy leads to augmented potential information density, meaning more watermark information can be embedded in a given amount of text without detection. During the watermark embedding process, there is a principle that the entropy should be distributed as evenly as possible across the entire sequence of watermarked tokens $T^N$ produced by the LLM, to avoid situations where entropy is high due to a single very low-probability token in the response.

Codable Watermarking~\cite{wangCodableWatermarkingInjecting2023b}
discusses the impact of the watermark message space size on text quality and computational complexity. Yoo et al.~\cite{yooRobustMultibitNatural2023}  analyze the impact of embedding a specific number of watermark bits per word (BPW) on the watermark performance. Qu et al.~\cite{quProvablyRobustMultibit2024} propose a watermarking method capable of extracting information for multiple payloads in linear time. Yoo et al.~\cite{yooAdvancingIdentificationMultibit2024} conduct ablation experiments on information density to explore the payload upper limit of the watermark algorithm.

However, maximizing information density must be balanced with other considerations. High information density may increase the visibility of the watermark to detection algorithms or human readers, thus compromising transparency. Moreover, overly dense watermarking changes may impact readability or alter meaning, which is counterproductive, especially in sensitive applications like legal documents or literary works. Hence, achieving an optimal information density in text watermarking necessitates a careful balance between embedding enough information to ensure the watermark's effectiveness and maintaining the original text's integrity and readability.

\subsection{Robustness}
Robustness refers to the capacity of a watermark to remain detectable in the face of various attacks. Watermarking technologies with greater robustness can ensure the continuity and coherence of information across a wider range of application scenarios. The level of robustness directly affects the feasibility and security of watermarking technologies. When evaluating robustness, the assessment involves exposing watermarked texts to various attacks (such as content modification, format conversion, model fine-tuning, etc.) and observing the persistence of the embedded watermark information. Metrics such as AUROC, F1 score, and Recall are used to measure the performance of watermark extraction and reconstruction under these attack conditions. Specifically, through the simulation of disruptive attacks on the watermarked text followed by attempting to recover the watermark via extraction techniques, the robustness is evaluated by comparing the success rate and watermark confidence before and after the attack. 

In addition to exploring the watermark robustness in various complex watermark attack scenarios~\cite{kirchenbauerReliabilityWatermarksLarge2023}, Zhao et al.~\cite{zhaoProtectingLanguageGeneration2023} investigate the impact of different decoding strategies on watermark performance by modifying the LLM's decoding strategies, such as beam-k and top-k. Meanwhile, Fernandez et al.~\cite{fernandezThreeBricksConsolidate2023} set different levels of watermark strength to evaluate text distortion under various watermark strengths as a measure of watermark robustness. However, these evaluations are specifically targeted at \textit{\textbf{Attack Robustness}} mentioned in Section~\ref{pd}. Presently, limited research on LLM watermarking has delved into the topic of \textit{\textbf{Security Robustness}}. Future studies should also focus on the watermark's own security robustness, that is, whether the watermark signal $W$ can be easily extracted and the watermark message $m$ inferred from it.

\subsection{Unforgeability}
Unforgeability focuses on the ability of watermarking technology to resist forgery or tampering, thereby ensuring the authenticity and reliability of watermark information. This objective can be achieved by training models to recognize specific watermark token distributions and then assessing the model's performance against both watermark spoofing attacks (attempts to forge watermark information) and watermark inference attacks (attempts to infer the watermarking strategy). The evaluation of unforgeability typically requires both qualitative analysis and quantitative testing, including metrics such as success rate and confidence levels.

When discussing the security of text watermarking, the primary aim is to prevent attackers from acquiring or cracking the watermark generation method. Watermark algorithms must exhibit a high level of unforgeability, making it challenging for attackers to identify their underlying generation logic. This involves the complexity of the algorithm or the use of mathematically hard-to-crack problems. 

In private detection scenarios, the imperceptibility of the watermark is crucial, meaning that the watermark's impact on the original content is difficult to detect. Research measures this attribute by testing the distinguishing ability of classifiers. Statistical methods might be used to analyze the embedding patterns of watermarks, requiring some understanding of the watermarking method. To enhance unforgeability, private detection scenarios should limit the frequency of detection. Sadasivan et al.~\cite{sadasivanCanAIGeneratedText2024} demonstrated that attackers would need more than a million queries to extract watermarks through privilege escalation potentially.

In public detection scenarios, where the detection algorithm is openly accessible, assessing the unforgeability of watermarks is more complex, as attackers can use this information to mount attacks. Ideal watermarking technology should ensure that even if attackers are aware of the algorithm details, they cannot successfully replicate or forge watermarks without the key. Extraction and reconstruction of the watermark algorithm should not leak detailed information about the generation method. Gu et al.~\cite{guLearnabilityWatermarksLanguage2024} utilize a model distillation approach to train a new model to learn the distribution of watermarked tokens. However, this forgery method is limited to algorithms that embed watermarks through modifications of logits. Pang et al.~\cite{pangAttackingLLMWatermarks2024a} argue that the deception attacks proposed in public detection settings could be generalized across all types of watermarks, requiring only a minimal number of queries to identify each token. Liu et al.~\cite{liuUnforgeablePubliclyVerifiable2024} are the first to assess the unforgeability properties of their watermarking algorithm formally and demonstrated that even attackers with access to watermark extraction and attempting to understand the watermark generation rules through an unlimited number of queries would find it difficult to deduce the watermark generation method. As for the publicly detectable watermarks referred to as~\cite{fairozePubliclyDetectableWatermarking2023}, the setup did not consider unforgeability measures, allowing the possibility of forging watermark messages without knowing the private key.

Assessing the unforgeability of these watermark algorithms necessitates the creation of complex attack algorithms, such as spoofing attacks. However, without knowledge of the generation architecture, it is challenging for attackers to succeed. Strong unforgeability implies that attackers find it difficult to infer the watermark generation method from the watermarked text, which imposes higher demands on the security and complexity of the watermarking algorithm, or necessitates the incorporation of robust cryptographic techniques.

\subsection{Cross-lingual Consistency}
Cross-lingual consistency is a critical measure to assess the efficacy of text watermarking when translated into other languages. The research~\cite{heCanWatermarksSurvive2024} aims to evaluate the consistency of current LLM watermarking algorithms across different languages, their performance in similar languages compared to distantly related languages, and the superiority of current semantic invariance-based watermarking methods over others.

Cross-lingual consistency is defined as the ability of a watermark embedded in text generated by LLMs to retain its strength after the text is translated into another language. Let the original strength of the watermark be denoted as a random variable $S$, and its strength after translation be denoted as $\hat{S}$. To quantitatively assess this consistency, the subsequent two metrics are employed:

1. Pearson Correlation Coefficient (PCC)

The Pearson Correlation Coefficient (PCC) is utilized to evaluate the linear correlation between $S$ and $\hat{S}$:

\begin{equation}
PCC(S, \hat{S}) = \frac{\text{cov}(S, \hat{S})}{\sigma_S \sigma_{\hat{S}}},
\end{equation}

\noindent where $\text{cov}(S, \hat{S})$ represents the covariance between $S$ and $\hat{S}$, and $\sigma_S$ and $\sigma_{\hat{S}}$ are the standard deviations of $S$ and $\hat{S}$, respectively. A PCC value close to 1 indicates a high degree of consistency in watermark strength trends across different languages.

2. Relative Error (RE)

In contrast to PCC, which captures the consistency of trends, the Relative Error (RE) is used to assess the magnitude of deviation between $S$ and $\hat{S}$:

\begin{equation}
RE(S, \hat{S}) = \mathbb{E}\left[\left|\frac{\hat{S} - S}{S}\right|\right] \times 100\%.
\end{equation}

A lower RE indicates that the watermark retains strength close to its original value after translation, signifying good cross-lingual consistency. To avoid instability when $S$ is close to 0, data is first aggregated by text length, and the original values of $S$ and $\hat{S}$ are replaced with the mean value of each group. Additionally, min-max normalization is applied to ensure all values are non-negative.

The consistency of watermark algorithms across different languages can be analyzed through two cross-linguistic consistency metrics: PCC and RE. He et al.~\cite{heCanWatermarksSurvive2024} have verified that current watermark algorithms struggle to maintain their advantages across different languages, with the robustness of watermarks being highly susceptible to cross-linguistic watermark removal attacks. This provides a new perspective on language translation for future watermark evaluation research.

\subsection{Radioactivity}
Radioactivity refers to the detectable traces left in a model when an LLM is fine-tuned using training data embedded with watermarks. These imprints serve as indicators that the output of the LLM could be used to fine-tune another model. This concept is vividly termed "radioactivity" in~\cite{sanderWatermarkingMakesLanguage2024a}, as it mirrors the way radioactive substances leave traceable residues in the environment. The radioactivity of LLM watermarks means that when watermark texts produced by LLM watermark algorithms are used as fine-tuning data, the characteristics of the watermark can still be preserved and radiated, allowing the watermark texts to be traced back through detection methods.

The strength of watermark radioactivity represents the ability to trace and source the watermarked data. Statistical tests based on cumulative scores and the number of tokens are typically used to determine whether text data is watermarked. Binomial or gamma distributions can be employed to calculate the probability (p-value) of obtaining a score higher than a certain threshold under the null hypothesis (i.e., the text is not watermarked). In radioactivity detection, a lower p-value (for example, less than $10^{-5}$ or $10^{-6}$) signifies a high level of confidence, indicating strong radioactivity has been detected. The detection of radioactivity can assist in tracing and sourcing the training data for LLMs.

\section{Future Directions}\label{sec12}
Although the previous sections thoroughly introduced the applications, theories, classifications, and evaluation systems of text watermarking, many challenges remain in this field. These include challenges related to rich information watermarking, asymmetric watermark encryption verification, and counteracting watermark attacks. The following sections will discuss these challenges in detail.
\subsection{Rich Information Watermarking Technology}
Despite the existence of numerous multi-bit LLM watermarking techniques, most research has been limited to algorithms involving a few watermark information bits, unable to embed rich information like traditional image~\cite{wanComprehensiveSurveyRobust2022} and audio~\cite{huaTwentyYearsDigital2016} watermarking technologies. Researchers should develop highly efficient large model watermark message embedding techniques to embed more hidden watermark information bits within the same text length. To achieve this, new algorithms must be explored and existing technologies optimized to increase the information embedding density in LLMs. This includes developing deep learning-based models that leverage the underlying hierarchical structure of the models for more efficient data encoding. Additionally, advanced compression techniques and information theory principles should be considered to reduce the required embedding space while maintaining or improving the robustness and transparency of the watermark. The development of such technologies can provide stronger tools for intellectual property protection and potentially open new applications in copyright management and data security. Ultimately, the goal is to achieve a text watermarking technology comparable to image and audio watermarking, capable of embedding large amounts of information without compromising the naturalness and readability of the text. This rich information watermarking scheme should minimize text quality loss due to watermark embedding and effectively resist text watermark attacks such as semantic rewriting, synonym replacement, and special symbol insertion.

\subsection{Asymmetric Encryption Verification Watermarking Technology}
Public trust and unforgeability are key factors in the widespread application of text watermarking technology. These technologies can only be effectively adopted when the public trusts the accuracy of watermark algorithms, which places high demands on the confidence levels of these algorithms. Fundamental steps to enhance trust include the comprehensive disclosure of watermark detection algorithms, enabling users to evaluate their principles and accuracy. Moreover, public trust can promote the development of academia and industry and can be strengthened through impartial evaluations by independent third-party platforms to reduce conflicts of interest. The formulation of government and regulatory guidelines is also an important way to ensure the fairness and transparency of watermarking technology and to enhance public trust.

To enhance the unforgeability of watermarking technology, it is necessary to introduce asymmetric encryption technology~\cite{bellareOptimalAsymmetricEncryption1995} into the watermark verification system of LLMs. As watermarking technology becomes more widespread, its verification process is increasingly open to the public, allowing anyone to check public watermarks to verify the legitimacy and integrity of data or models. However, to prevent attacks at various stages of watermarking, some key watermark encryption keys must be restricted by permissions. In the research of AI watermark encryption verification technology, exploring the application of asymmetric encryption technology in the extraction and verification process of LLM watermarks is particularly important. The future may witness further enhancement of technology security through the adoption of digital signature technology~\cite{merkleDigitalSignatureBased1988} and the use of one-way pseudo-random functions~\cite{impagliazzoPseudorandomGenerationOneway1989} to guide the generation of watermarks. An important direction for the future is the development of a distributed public watermark verification system based on a shared key database. Such a system can effectively resist attacks that may occur during the watermark verification stage. By introducing a verification mechanism involving multiple parties, the system's security can be enhanced, and the accuracy and reliability of watermark detection can be improved. This system design aims to build a more robust and trustworthy watermark technology framework, ensuring the legitimacy and integrity of data and models are widely and effectively protected.

\subsection{Countering Watermark Attacking Technology}
Although experimental evidence~\cite{sadasivanCanAIGeneratedText2024} suggests that text-based watermarks can be designed to resist various attacks such as paraphrasing and copy-and-paste, their robustness is a function of text length. Text fragments under 1,000 words become more challenging, with efficiency steadily declining as text size decreases. We believe that two feasible actions need to be taken simultaneously to form a comprehensive approach to countering watermark attacks. First, robust watermarking techniques should be researched to withstand various attacks and processes, such as watermark removal and theft, applicable for copyright protection and model usage tracking. Second, a fragile watermark scheme for LLMs should be proposed, similar to fragile watermarks in image watermarking. Fragile watermarks are highly sensitive to modifications. If data or models are tampered with, the fragile watermark will be destroyed. They are suitable for content authentication and integrity verification.

Robust and fragile watermarks should be used simultaneously in data or models. Robust watermarks are used for long-term tracking and rights protection, while fragile watermarks are used for immediate tamper detection and integrity verification. This combined approach can create a comprehensive watermark system that protects individual entities' private rights while meeting public regulatory needs. By applying both robust and fragile watermarks, copyright protection for data and models can be maintained over the long term, and the integrity and authenticity of data and models can be effectively monitored and verified. This design helps establish a secure, transparent, and sustainable digital content and service ecosystem.

\subsection{Integration of Other Identification Technologies}
The future direction of LLM watermarking for identification can integrate various advanced identification technologies to enhance the security and reliability of LLM identity recognition systems. Key technologies to consider include behavior classification~\cite{zhong2012keystroke} (analyzing user behavior patterns such as typing speed and interaction habits); deep fake detection~\cite{zhao2021multi} (identifying AI-generated content); and biometric methods~\cite{jain2006biometrics} (such as fingerprint, facial, iris, and voice recognition). Additionally, multimodal recognition~\cite{alay2020deep} (combining multiple data sources like visual, audio, and textual information) and intelligent identity verification~\cite{mohammed2013intelligent} (utilizing artificial intelligence and machine learning for dynamic analysis) can be integrated. By combining these advanced identification technologies with LLM watermarking and adapting these technologies to the characteristics of AIGC and LLM application scenarios, we can develop a more secure, accurate, and adaptive LLM identity recognition system capable of addressing evolving security threats and improving user experience.

\section{Conclusion}\label{sec13}
In this review, we comprehensively explore the developments and significance of constructing LLM identification systems using LLM watermarking technology. With the widespread application of LLMs, ensuring the distinguishability, unforgeability, and traceability of LLM behavior has become particularly critical.  As the LLM application system evolves towards a multi-centric system, the positions of various participants become balanced. To ensure trustworthy collaboration and minimize suspicion among participants, LLM identity recognition via watermarking can be employed to ensure identification and behavior traceability throughout the LLM lifecycle.

To advance the development of LLM watermarking technology, we have undertaken several initiatives.
We have established a mathematical framework centered on information theory, providing a solid theoretical foundation for the research and optimization of watermarking techniques. Through detailed classification, mathematical description, and comprehensive evaluation metrics for watermarking technology, we can reflect the preferences of different participants and promote the development of LLM watermarking technology. This systematic classification and elucidation not only facilitate comparison and evaluation between methods but also enhance the understanding of watermarking technology's security, offering new research and evaluation directions.

We anticipate that future research will delve deeper into the development of more efficient and secure identification technologies, particularly in the domains of rich information watermarking, asymmetric watermark encryption verification, countering watermark attacks, and integration of other identification technologies to meet increasingly complex application demands and security challenges in the field of intelligence identification.

\newpage

\bibliography{reference}

\end{document}